\documentclass[12pt]{article}
\usepackage[dvips]{graphicx}
\usepackage{epsfig}
\usepackage{latexsym,amsmath,amsfonts,amssymb}
\usepackage[latin1]{inputenc}
\usepackage[american]{babel}
\usepackage[dvips]{graphicx}
\usepackage{bbm}
\def\prl{Phys. Rev. Lett.}
\def\pr{Phys. Rev.}

\def\im{Invent. Math.}

\newcommand{\be}{\begin{equation}}
\newcommand{\ee}{\end{equation}}
\newcommand{\beq}{\begin{equation}}
\newcommand{\eeq}{\end{equation}}
\newcommand{\bea}{\begin{eqnarray}}
\newcommand{\eea}{\end{eqnarray}}
\newcommand{\nn}{\nonumber}

\newcommand{\ba}{\begin{eqnarray}}
\newcommand{\ea}{\end{eqnarray}}
\newcommand{\bal}{\begin{aligned}}
\newcommand{\eal}{\end{aligned}}

\begin{document}
\baselineskip=15.5pt
\pagestyle{plain}
\setcounter{page}{1}
%--------+---------+---------+---------+---------+---------+---------+
%Body

% Ofer's definitions

\def\del{{\partial}}
\def\vev#1{\left\langle #1 \right\rangle}
\def\cn{{\cal N}}
\def\co{{\cal O}}
%\newfont{\Bbb}{msbm10 scaled 1200}     %instead of eusb10
%\newcommand{\mathbb}[1]{\mbox{\Bbb #1}}
\def\IC{{\mathbb C}}
\def\IR{{\mathbb R}}
\def\IZ{{\mathbb Z}}
\def\RP{{\bf RP}}
\def\CP{{\bf CP}}
\def\Poincare{{Poincar\'e }}
\def\tr{{\rm tr}}
\def\tp{{\tilde \Phi}}

\def\TL{\hfil$\displaystyle{##}$}
\def\TR{$\displaystyle{{}##}$\hfil}
\def\TC{\hfil$\displaystyle{##}$\hfil}
\def\TT{\hbox{##}}
\def\HLINE{\noalign{\vskip1\jot}\hline\noalign{\vskip1\jot}}
\def\seqalign#1#2{\vcenter{\openup1\jot
   \halign{\strut #1\cr #2 \cr}}}
\def\lbldef#1#2{\expandafter\gdef\csname #1\endcsname {#2}}
\def\eqn#1#2{\lbldef{#1}{(\ref{#1})}%
\begin{equation} #2 \label{#1} \end{equation}}
\def\eqalign#1{\vcenter{\openup1\jot
     \halign{\strut\span\TL & \span\TR\cr #1 \cr
    }}}
\def\eno#1{(\ref{#1})}
\def\href#1#2{#2}
\def\half{{1 \over 2}}

%--------+---------+---------+---------+---------+---------+---------+
%Hirosi's macros:

\def\ads{{\it AdS}}
\def\adsp{{\it AdS}$_{p+2}$}
\def\cft{{\it CFT}}

\newcommand{\ber}{\begin{eqnarray}}
\newcommand{\eer}{\end{eqnarray}}

\newcommand{\beqar}{\begin{eqnarray}}
\newcommand{\cN}{{\cal N}}
\newcommand{\cO}{{\cal O}}
\newcommand{\cA}{{\cal A}}
\newcommand{\cT}{{\cal T}}
\newcommand{\cF}{{\cal F}}
\newcommand{\cC}{{\cal C}}
\newcommand{\cR}{{\cal R}}
\newcommand{\cW}{{\cal W}}
\newcommand{\cK}{{\cal K}}
\newcommand{\eeqar}{\end{eqnarray}}
\newcommand{\tht}{\thteta}
\newcommand{\lm}{\lambda}\newcommand{\Lm}{\Lambda}
\newcommand{\eps}{\epsilon}

%--------+---------+---------+---------+---------+---------+---------+

\newcommand{\nonu}{\nonumber}
\newcommand{\oh}{\displaystyle{\frac{1}{2}}}
\newcommand{\dsl}
   {\kern.06em\hbox{\raise.15ex\hbox{$/$}\kern-.56em\hbox{$\partial$}}}
\newcommand{\id}{i\!\!\not\!\partial}
\newcommand{\as}{\not\!\! A}
\newcommand{\ps}{\not\! p}
\newcommand{\ks}{\not\! k}
\newcommand{\D}{{\cal{D}}}
\newcommand{\dv}{d^2x}
\newcommand{\Z}{{\cal Z}}
\newcommand{\N}{{\cal N}}
\newcommand{\Dsl}{\not\!\! D}
\newcommand{\Bsl}{\not\!\! B}
\newcommand{\Psl}{\not\!\! P}
\newcommand{\eeqarr}{\end{eqnarray}}
\newcommand{\ZZ}{{\rm \kern 0.275em Z \kern -0.92em Z}\;}
%--------------------------------Alfonso's definitions%%%%%%%%%%%%%

% DEFINITIONS

\def\del{{\delta^{\hbox{\sevenrm B}}}} \def\ex{{\hbox{\rm e}}}
\def\azb{A_{\bar z}} \def\az{A_z} \def\bzb{B_{\bar z}} \def\bz{B_z}
\def\czb{C_{\bar z}} \def\cz{C_z} \def\dzb{D_{\bar z}} \def\dz{D_z}
\def\im{{\hbox{\rm Im}}} \def\mod{{\hbox{\rm mod}}} \def\tr{{\hbox{\rm Tr}}}
\def\ch{{\hbox{\rm ch}}} \def\imp{{\hbox{\sevenrm Im}}}
\def\trp{{\hbox{\sevenrm Tr}}} 
\def\rl{\Lambda_{\hbox{\sevenrm R}}} \def\wl{\Lambda_{\hbox{\sevenrm W}}}
\def\fc{{\cal F}_{k+\cox}} \def\vev{vacuum expectation value}
\def\nodiv{\mid{\hbox{\hskip-7.8pt/}}}
\def\ie{{\em i.e.}}
\def\ie{\hbox{\it i.e.}}

\def\CC{{\mathchoice
{\rm C\mkern-8mu\vrule height1.45ex depth-.05ex
width.05em\mkern9mu\kern-.05em}
{\rm C\mkern-8mu\vrule height1.45ex depth-.05ex
width.05em\mkern9mu\kern-.05em}
{\rm C\mkern-8mu\vrule height1ex depth-.07ex
width.035em\mkern9mu\kern-.035em}
{\rm C\mkern-8mu\vrule height.65ex depth-.1ex
width.025em\mkern8mu\kern-.025em}}}

\def\RR{{\rm I\kern-1.6pt {\rm R}}}
\def\NN{{\rm I\!N}}
\def\ZZ{{\rm Z}\kern-3.8pt {\rm Z} \kern2pt}
\def\IB{\relax{\rm I\kern-.18em B}}
\def\ID{\relax{\rm I\kern-.18em D}}
\def\II{\relax{\rm I\kern-.18em I}}
\def\IP{\relax{\rm I\kern-.18em P}}
\newcommand{\CS}{{\scriptstyle {\rm CS}}}
\newcommand{\CSs}{{\scriptscriptstyle {\rm CS}}}
\newcommand{\rc}{\nonumber\\}
\newcommand{\bear}{\begin{eqnarray}}
\newcommand{\eear}{\end{eqnarray}}
\newcommand{\W}{{\cal W}}
\newcommand{\F}{{\cal F}}
\newcommand{\x}{{\cal O}}
\newcommand{\LL}{{\cal L}}

\def\mani{{\cal M}}
\def\calo{{\cal O}}
\def\calb{{\cal B}}
\def\calw{{\cal W}}
\def\calz{{\cal Z}}
\def\cald{{\cal D}}
\def\calc{{\cal C}}
\def\to{\rightarrow}
\def\ele{{\hbox{\sevenrm L}}}
\def\ere{{\hbox{\sevenrm R}}}
\def\zb{{\bar z}}
\def\wb{{\bar w}}
\def\nodiv{\mid{\hbox{\hskip-7.8pt/}}}
\def\menos{\hbox{\hskip-2.9pt}}
\def\dr{\dot R_}
\def\drr{\dot r_}
\def\ds{\dot s_}
\def\da{\dot A_}
\def\dga{\dot \gamma_}
\def\ga{\gamma_}
\def\dal{\dot\alpha_}
\def\al{\alpha_}
\def\cl{{closed}}
\def\cls{{closing}}
\def\vev{vacuum expectation value}
\def\tr{{\rm Tr}}
\def\to{\rightarrow}
\def\too{\longrightarrow}

% Umut likes:

\def\a{\alpha}
\def\b{\beta}
\def\c{\gamma}
\def\d{\delta}
\def\e{\epsilon}           % Also, \varepsilon
\def\f{\phi}               %      \varphi
\def\vf{\varphi}  \def\tvf{\tilde{\varphi}}
\def\vp{\varphi}
\def\g{\gamma}
\def\h{\eta}
\def\j{\psi}
\def\k{\kappa}                    % Also, \varkappa (see below)
\def\l{\lambda}
\def\m{\mu}
\def\n{\nu}
\def\o{\omega}  \def\w{\omega}
\def\p{\pi}
\def\q{\theta}  \def\th{\theta}                  %     \vartheta
\def\r{\rho}                                     %     \varrho
\def\s{\sigma}                                   %     \varsigma
\def\t{\tau}
\def\u{\upsilon}
\def\x{\xi}
\def\z{\zeta}
\def\pt{\tilde{\varphi}}
\def\tt{\tilde{\theta}}
\def\lab{\label}
\def\6{\partial}
\def\wg{\wedge}
\def\atanh{{\rm arctanh}}
\def\bpsi{\bar{\psi}}
\def\bt{\bar{\theta}}
\def\bvf{\bar{\varphi}}

%%%
%%%%%%% Definitions for the current file
\newcommand{\td}{d}
\newcommand{\vol}[1]{\textrm{Vol(}#1\textrm{)}}
\newcommand{\Thq}{\Theta\left(\r-\r_q\right)}
\newcommand{\Dq}{\d\left(\r-\r_q\right)}
\newcommand{\kten}{\kappa^2_{\left(10\right)}}
\newcommand{\pbi}[1]{\imath^*\left(#1\right)}
\newcommand{\ho}{\hat{\omega}}
\newcommand{\hth}{\hat{\th}}
\newcommand{\hf}{\hat{\f}}
\newcommand{\hj}{\hat{\j}}

%
% FONTS

%\newfont{\headfont}{cmbx10 scaled 1440}
\newfont{\namefont}{cmr10}
%\newfont{\initialfont}{cmr10 scaled 1200}
\newfont{\addfont}{cmti7 scaled 1440}
\newfont{\boldmathfont}{cmbx10}
%\newfont{\figfont}{cmr7 scaled 1200}
\newfont{\headfontb}{cmbx10 scaled 1728}
%%%%%%%%%%%%%%%%%%%%%%%%%%%%%%%%%%%%%%%%%%%%%%%%%%%%%%%%%%%%%%%%%%%%%%%%%%%%
%%%%%%%%%%%%%%%Stefano and Francesco fonts%%%%%%%%%%%%%%%%%%%%%%%%%%%%%%%%
%%%%%%%%%%%%%%%%%%%%%%%%%%%%%%%%%%%%%%%%%%%%%%%%%%%%%%%%%%%%%%%%%%%%%%%%%%%%
\newcommand{\re}{\,\mathbb{R}\mbox{e}\,}
\newcommand{\hyph}[1]{$#1$\nobreakdash-\hspace{0pt}}
\providecommand{\abs}[1]{\lvert#1\rvert}
\newcommand{\Nugual}[1]{$\mathcal{N}= #1 $}
\newcommand{\sub}[2]{#1_\text{#2}}
\newcommand{\partfrac}[2]{\frac{\partial #1}{\partial #2}}
\newcommand{\bsp}[1]{\begin{equation} \begin{split} #1 \end{split} \end{equation}}
\newcommand{\calF}{\mathcal{F}}
\newcommand{\calO}{\mathcal{O}}
\newcommand{\calM}{\mathcal{M}}
\newcommand{\calV}{\mathcal{V}}
\newcommand{\bbZ}{\mathbb{Z}}
\newcommand{\bbC}{\mathbb{C}}

\numberwithin{equation}{section}

\newcommand{\Tr}{\mbox{Tr}}    % trace over gauge indices

%%%%%%%%%%%%%%%%%%%%%%%%%%%%%%%%%%%%%%%%%%%%%%%%%%%%%%%%%%%%%%%%%%%%%%%%%%%%
%%%%%%%%%%%%%%%%%%%%%%%%%%%%%%%%%%%%%%%%%%%%%%%%%%%%%%%%%%%%%%%%%%%%%%%%%%%%

%
\renewcommand{\theequation}{{\rm\thesection.\arabic{equation}}}
\begin{titlepage}
\rightline{SISSA 22/2010/EP}  
\vspace{0.1in}

\begin{center}
\Large \bf Holographic duals of  SQCD models in low dimensions 
\end{center}
\vskip 0.2truein
\begin{center}
Daniel Are\'an ${}^{\dagger}$\footnote{arean@sissa.it}, 
Eduardo Conde  ${}^{*}$\footnote{eduardo@fpaxp1.usc.es}, 
Alfonso V. Ramallo${}^{*}$\footnote{alfonso@fpaxp1.usc.es} and Dimitrios Zoakos${}^{*\,\ddagger}$\footnote{dimitrios.zoakos@fc.up.pt} \\
\vspace{0.2in}
${}^{*}$\it{
Departamento de  F\'{\i}sica de Part\'{\i}culas, Universidade
de Santiago de
Compostela\\and\\Instituto Galego de F\'{\i}sica de Altas
Enerx\'{\i}as (IGFAE)\\E-15782, Santiago de Compostela, Spain
}
\\
\vspace{0.2in}
${}^{\dagger}$  \it{ SISSA and INFN-Sezione di Trieste,\\ Via Bonomea 265, 34136 Trieste, Italy}

\vspace{0.2in}
${}^{\ddagger}$ 
{\it
Departamento de F\'{\i}sica e Astronomia \& Centro de F\'{\i}sica do Porto,\\
Faculdade de Ci\^{e}ncias da Universidade do Porto,\\
Rua do Campo Alegre 687, 4169--007 Porto, Portugal}

\vspace{0.2in}
\end{center}
\vspace{0.2in}
%\begin{center}
\centerline{{\bf Abstract}}
We obtain gravity duals to supersymmetric gauge theories in  two and three spacetime dimensions with unquenched flavor. The supergravity solutions are generated by a set of color branes wrapping a compact cycle in a Calabi-Yau threefold, together with another set of flavor branes extended along the directions orthogonal to the cycle wrapped by the color branes. We construct supergravity backgrounds which include the backreaction induced by a smeared set of flavor branes, which act as delocalized dynamical sources of the different supergravity fields.

\smallskip
\end{titlepage}
\setcounter{footnote}{0}

\tableofcontents

\section{Introduction}
The AdS/CFT correspondence \cite{Maldacena:1997re, Gubser:1998bc,Witten:1998qj, reviewsAdSCFT} is one of the major achievements of theoretical physics of the last few years. Indeed, this correspondence has provided analytic tools to explore the strong coupling regime of a large variety of gauge theories in the planar limit $N_c\to\infty$. Originally, the AdS/CFT correspondence was formulated as a duality between ${\cal N}=4$ super Yang-Mills (SYM) gauge theories in four-dimensions and type IIB supergravity in  $AdS_5\times S^5$. Extending the duality beyond this highly symmetric theory is clearly a problem which deserves to be studied and, for this reason, there has been much effort devoted to constructing supergravity duals to theories with lower amounts of supersymmetry and different field content in several spacetime dimensions.

A general approach to construct gravity duals with lower amounts of supersymmetry consists of considering higher dimensional branes wrapping cycles inside a non-compact manifold of special holonomy \cite{Maldacena:2000mw}. At energies small compared with the size of the cycles these solutions provide gravity duals of SYM theories living on the unwrapped dimensions of the brane. A very useful strategy to find supergravity solutions corresponding to wrapped branes is the use of the appropriate gauged supergravities, which are lower-dimensional theories in which the brane behaves as a domain wall. In these theories one can identify the spin connection along the wrapped cycle with a gauge field, implementing in this way the so-called topological twist \cite{Bershadsky:1995qy}. In this approach the ten-dimensional background is obtained from the lower dimensional one by using the corresponding uplifting formulae. This program has been successfully carried out to find duals to gauge theories in several spacetime dimensions with different amounts of supersymmetry \cite{CV}-\cite{Arean:2009gc}. 

Another important generalization of the AdS/CFT  correspondence has been the addition  of matter degrees of freedom transforming in the fundamental representation of the gauge group,  \ie\ quarks. This new matter sector can be added by including flavor branes, which extend along all gauge theory directions and, in order to make its worldvolume symmetry a global symmetry from the gauge theory point of view, they should be also  extended along some other non-compact directions \cite{KK}. When the number $N_f$ of flavor branes is small compared with the number $N_c$ of colors, one can treat the flavor branes as probes which do not modify the background created by the color branes. This defines the so-called quenched approximation which, on the field theory side, corresponds to considering the quarks as external fields that do not run in loops and, thus, to neglect the quantum effects produced by the fundamentals. By studying the worldvolume physics in this probe approximation one can address many interesting problems such as, for example, the meson spectrum \cite{Kruczenski:2003be} (see \cite{Erdmenger:2007cm} for a review). 

When the number of flavors is of the order of the number of colors ($N_f\sim N_c$) one must face the problem of studying the deformation produced by the flavor branes on the geometry. Computing this backreaction is, in general, a very complicated task. Indeed, in these unquenched setups the flavor branes should be regarded as dynamical sources of the different supergravity fields and, unless one adopts some simplifying assumptions, the problem becomes, in many cases, intractable.   For this reason, in this paper we will consider the case in which the flavor branes are homogeneously distributed by forming a smeared set. This smearing technique to add unquenched flavor was first introduced in \cite{Bigazzi:2005md} in  a non-critical string framework and in \cite{CNP} in a well-controlled ten-dimensional context. Subsequently, this approach has been successfully applied in several brane setups \cite{Angel}-\cite{Bigazzi:2009bk} (see \cite{Nunez:2010sf} for a review). 

In this paper we study the addition of unquenched flavor to backgrounds constructed by wrapping D3-, D4- and D5-branes along two- and three-cycles of a Calabi-Yau threefold. The unflavored backgrounds, which are dual to gauge theories in two and three dimensions, have been constructed in refs. \cite{Maldacena:2000mw,Gomis:2001aa,Gauntlett:2001ur} from the appropriate gauged supergravity. Here we reformulate these models in terms of a new set of variables which greatly simplify their interpretation and allows the introduction of the corresponding flavor deformation in a neat way. After this reformulation the background is determined by a set of functions depending on two variables, which represent the radial variables inside the Calabi-Yau cone and in the space transverse to it. These functions satisfy a set of first-order differential equations, which can be recast compactly in terms of a generalized calibration form. 

The flavor branes for the setups considered here are extended across the directions normal to the cycle that the color branes wrap. In order to obtain the backreaction of these flavor branes on the background the key point is realizing that they act as sources for the RR field and, as a consequence, they induce a violation of the corresponding Bianchi identity. For a set of delocalized flavor branes the violation of the Bianchi identity compatible with the preserved supersymmetry can be encoded in a simple modification (parametrized by two functions) of the ansatz of the RR form, while the ten-dimensional metric has the same form as in the unflavored case. By requiring certain regularity conditions, the functions which parametrize the flavor deformation can be greatly constrained and, in fact, they can be exactly matched with those obtained by a careful microscopic counting of the RR charge distribution produced by a family of supersymmetric embeddings. In this paper  we will determine these embeddings and we shall perform the matching with the macroscopic analysis for the three cases analyzed. Moreover, the backreacted backgrounds will be found by a numerical integration of a PDE with sources. 

The rest of this paper is organized as follows. In section \ref{D3-section} we will consider our first background, namely the one generated by D3-branes wrapping a two-cycle in a Calabi-Yau threefold and preserving four supercharges. This background is dual to a two-dimensional gauge theory with ${\cal N}=(2,2)$ supersymmetry. We first reformulate the results of \cite{Maldacena:2000mw} for the unflavored case by introducing a second-order master differential equation whose solutions determine the unflavored background in implicit form. Then, we analyze the flavor deformation induced by a smeared set of D3-branes extended along two of the directions normal to the cycle within the Calabi-Yau cone. This deformation is analyzed first from the macroscopic point of view and, then, these results are reproduced from a microscopic analysis of the embeddings. Section  \ref{D3-section} ends with the presentation of the numerical results for the flavored model.

Section \ref{D4-section} is devoted to the study of a gravity dual to ${\cal N}=2$ gauge theories in three-dimensions generated by wrapping D4-branes on a two-cycle of a Calabi-Yau cone of complex dimension three. The analysis in this section runs completely in parallel with the one performed in section \ref{D3-section}. In section 
 \ref{D5-section} we consider the backgrounds with four supersymmetries which are generated by D5-branes wrapping a three-cycle. The corresponding unflavored solution was found in refs. \cite{Gomis:2001aa,Gauntlett:2001ur}, while the macroscopic approach to adding flavor was first introduced in ref. \cite{Gaillard:2008wt}. In  section  \ref{D5-section} we identify the deformation which corresponds to a regular charge density distribution of flavor branes and we find a family of embeddings of D5-branes which reproduces the result of the macroscopic formalism. Finally, in section \ref{conclusions} we summarize our results and present our conclusions.

\section{${\cal N}=(2,2)$ 2d SQCD from wrapped D3-branes}
\label{D3-section}
The background dual to the two-dimensional gauge theory which we will analyze in this section is generated by a stack of $N_c$ D3-branes wrapping a two-cycle of a Calabi-Yau cone of complex dimension three, according to the setup:
\begin{center}
\begin{tabular}{|c|c|c|c|c|c|c|c|c|c|c|}
\multicolumn{3}{c}{ }&
\multicolumn{6}{c}
{$\overbrace{\phantom{\qquad\qquad\qquad\qquad}}^{\text{CY}_3}$}\\
\hline
&\multicolumn{2}{|c|}{$\mathbb{R}^{1,1}$}
&\multicolumn{2}{|c|}{$S^2$}
&\multicolumn{4}{|c|}{$N_4$}
&\multicolumn{2}{|c|}{$\mathbb{R}^{2}$}\\
\hline
$N_c$ D$3$ &$-$&$-$&$\bigcirc$&$\bigcirc$&$\cdot$&$\cdot$
&$\cdot$&$\cdot$&$\cdot$&$\cdot$\\
\hline
\end{tabular}
\end{center}
where $S^2$ represents the directions of the two-cycle ${\cal C}_2$ that the branes are wrapping and $N_4$ are the directions of the normal bundle to ${\cal C}_2$. In the above setup a circle represents a wrapped direction, whereas the symbols ``-" and ``." denote unwrapped worldvolume and transverse directions respectively. We shall parametrize the cycle ${\cal C}_2$ by means of two angular coordinates $(\theta, \phi)$ with $0\le \theta<\pi$ and $0\le \phi <2\pi$. The normal bundle $N_4$ is non-compact and will be parametrized by a radial coordinate $\sigma$, together with  three  other angular coordinates. Moreover, the transverse space $\mathbb{R}^{2}$ will be described by polar coordinates $(\rho, \lambda)$ with $0\le \rho<\infty$ and $0\le \lambda<2\pi$. The ansatz for the ten-dimensional metric of type IIB supergravity corresponding to this setup is:
\bear 
&&ds^2_{10}\,=\,H^{-{1\over 2}}\,
\Big[\,dx^{2}_{1,1}\,+\,{z\over m^2}\,\big(\,(d\theta)^2\,+\,\sin^2\theta\, (d\phi)^2\,\big)
\,\Big]\,+\,\rc\rc
&&
+\,H^{{1\over 2}}\,\Big[\,{1\over z^{{1\over 2}}}\,\Big(\,(d\sigma)^2\,+\,
{\sigma^2\over 4}\,
\big[\,(w^1)^2\,+\,(w^2)^2\,+\, (w^3+\cos\theta\, d\phi)^2\,\big]\,\Big)
+\,
(d\rho)^2\,+\,\rho^2\,(d\lambda)^2\,\Big]\,\,,\qquad\qquad
\label{D3-metric}
\eear
where  $dx^{2}_{1,1}$ denotes the Minkowski metric in $1+1$ dimensions, 
the $w^i$  $(i=1,2,3)$ are SU(2) left-invariant one-forms satisfying $dw^i={1\over 2}\,\epsilon^{ijk}\,w^j\wedge w^k$ and $m$ is a constant with units of mass which, for convenience we will take as:
\beq
{1\over m^2}\,=\,\sqrt{4\pi g_s\,N_c}\,\alpha'\,\,,
\label{m-D3}
\eeq
with $g_s$ and $\alpha'$ being respectively the string coupling constant and the Regge slope of superstring theory. The  metric ansatz (\ref{D3-metric}) contains two functions, 
$z$, which controls the size of the cycle ${\cal C}_2$, and the warp factor $H$. Both of them should be considered as functions of the two radial coordinates $\rho$ and $\sigma$, \ie\ $z=z(\rho,\sigma)$, $H=H(\rho, \sigma)$. Moreover, as any other background generated by D3-branes, the ansatz should be endowed with a self-dual  RR five-form $F_5$. Let us write $F_5$ as:
\beq
F_5\,=\,{\cal F}_5\,+\,{}^*{\cal F}_5\,\,,
\label{F5-total}
\eeq
where ${\cal F}_5$ can be represented in terms of a four-form potential ${\cal C}_4$ as:
\beq
{\cal F}_5\,=\,d\,{\cal C}_4\,\,.
\label{F5-dC4}
\eeq
The explicit ansatz that we will adopt  for ${\cal C}_4$  is:
\beq
{\cal C}_4\,=\,g_1\,w^1\wedge w^2\wedge (w^3+\cos\theta d\phi)\wedge d\lambda\,+\,
g_2\,\Omega_2\wedge (w^3+\cos\theta d\phi)\wedge d\lambda\,\,,
\label{C4}
\eeq
with  $g_1$ and $g_2$  being two functions depending on both radial coordinates $\rho$ and $\sigma$ and $\Omega_2$ being the volume form of the two-sphere:
\beq
\Omega_2\,=\,\sin\theta d\theta\wedge d\phi\,\,.
\label{Omega2}
\eeq

We will determine the functions $z$, $H$, $g_1$ and $g_2$ of our ansatz by imposing the condition that our background preserves four supersymmetries. In order to specify this condition it is convenient to fix the following vielbein basis of one-forms for the metric (\ref{D3-metric}):
\bear
&&e^{0,1}\,=\,H^{-{1\over 4}}\,dx^{0,1}\,\,,\qquad\qquad
e^2\,=\,{H^{-{1\over 4}}\over m}\,\,\sqrt{z}\,d\theta\,\,,\qquad\qquad
e^3\,=\,{H^{-{1\over 4}}\over m}\,\,\sqrt{z}\,\sin\theta\,d\phi\,\,,\rc\rc
&&e^{4}\,=\,{H^{{1\over 4}}\over z^{{1\over 4}}}\,d\sigma\,\,,\qquad\qquad
\qquad
e^{5}\,=\,{H^{{1\over 4}}\over 2 z^{{1\over 4}}}\,\,\sigma\, w^1\,\,,
\qquad\qquad\quad
e^{6}\,=\,{H^{{1\over 4}}\over 2 z^{{1\over 4}}}\,\,\sigma\, w^2\,\,,\rc\rc
&&e^{7}\,=\,{H^{{1\over 4}}\over 2 z^{{1\over 4}}}\,\,\sigma\, 
(w^3+\cos\theta d\phi)\,\,,\qquad\qquad
e^{8}\,=\,H^{{1\over 4}}\,d\rho\,\,,\qquad\qquad
e^{9}\,=\,H^{{1\over 4}}\,\rho\,d\lambda\,\,.
\label{10-D3-frame}
\eear
We shall now impose the vanishing of the supersymmetry variations of the dilatino and gravitino of the the type IIB theory. The corresponding  Killing spinors $\epsilon$ are required to satisfy the following set of projections:
\beq
\Gamma_{2356}\,\epsilon\,=\,\epsilon\,\,,\qquad\qquad
\Gamma_{5647}\,\epsilon\,=\,-\epsilon\,\,,\qquad\qquad
\Gamma_{0123}\,(\,i\tau_2)\,\epsilon\,=\,\epsilon\,\,,
\label{D3-projections}
\eeq
where $\epsilon$ is a doublet of Majorana-Weyl spinors of fixed ten-dimensional chirality and $\tau_2$ is the second Pauli matrix, which acts on the doublet $\epsilon$. In 
(\ref{D3-projections}) $\Gamma_{a_1\,a_2\,\cdots}$ are antisymmetrized products of constant Dirac matrices in the vielbein basis (\ref{10-D3-frame}). Moreover, the four unknown functions $z$, $H$, $g_1$ and $g_2$ must satisfy the following system of first-order differential equations:
\bear
&&z'\,=\,8m^2\,{\sqrt{z}\over \rho\sigma^2}\,(\, g_2\,-\,g_1\,)\,\,,
\qquad\qquad\qquad
\dot z\,=\,{m^2\sigma \over \sqrt{z}}\, H\,\,,\rc\rc
&&g'_1\,=\,{1\over 8}\,{\rho\sigma^3\over \sqrt{z}}\,\,\dot H\,-\,{m^2\over 16}\,
{\rho\sigma^4\over z^2}\,\,H^2\,\,,
\qquad\qquad
\dot g_1\,=\,-{1\over 8}\,{\rho\sigma^3\over  z}\,\,H'\,+\,
m^2\,{\sigma H\over z^{{3\over 2}}}\,(\,g_2\,-\,g_1\,)\,\,,\rc\rc
&& g_2'\,=\,-{1\over 4}\,{\rho\sigma^2\over \sqrt{z}}\,H\,\,,
\qquad\qquad\qquad\qquad\qquad
\dot{ g_2}\,=\,{2\over \sigma}\,(\, g_2\,-\,g_1\,)\,\,,
\label{BPS-D3}
\eear
where the prime (dot) denotes the partial derivative with respect to $\rho$ ($\sigma$). Similarly to other backgrounds studied in \cite{Angel, Arean:2008az, Ramallo:2008ew, Arean:2009gc}, the system (\ref{BPS-D3}) can be reduced to the following PDE for the function $z(\rho,\sigma)$:
\beq
2\rho\,z\,\sqrt{z}\,\big(\,\sigma\,\ddot z\,+\,\dot z\,\big)\,=\,\sigma\,\big(\,\rho\,z'^2\,
-\,2z\,z'\,-\,2\rho z\,z''\,\big)\,\,.
\label{PD3-D3unflavored}
\eeq
One can verify that, if $z(\rho,\sigma)$ is known, the other three functions $H$, $g_1$ and $g_2$ of our ansatz can be determined from the BPS system (\ref{BPS-D3}). Moreover, if (\ref{BPS-D3}) holds, the second-order equations of motion of type IIB supergravity are satisfied. Notice also that the Killing spinors $\epsilon$ satisfy  the three independent projections (\ref{D3-projections}), which means that the background is ${1\over 8}$-supersymmetric and four supersymmetries are unbroken. It is easy to check that there are two supercharges of each two-dimensional chirality, as it corresponds to the gravity dual of an ${\cal N}=(2,2)$ gauge theory. 

The BPS system (\ref{BPS-D3}) can be written in a very compact and suggestive form in terms of the so-called (generalized) calibration form ${\cal K}$. For the case at hand ${\cal K}$ is a four-form, which can be represented in the vielbein basis (\ref{10-D3-frame}) as:
\beq
{\cal K}\,=\,{1\over 4!}\,{\cal K}_{a_1\cdots a_4}\,e^{a_1\,\cdots a_4}\,\,,
\eeq
where $e^{a_1\,\cdots a_4}=e^{a_1}\wedge \cdots \wedge e^{a_4}$. The different components of ${\cal K}$ are obtained from the fermionic bilinears:
\beq
{\cal K}_{a_1\cdots a_4}\,=\,H^{{1\over 4}}\,\,\epsilon^{\dagger}\,i\tau_2\,\Gamma_{a_1\,\cdots a_4}\,\epsilon\,\,,
\label{K-bilinear-D3}
\eeq
where $\epsilon$ is a Killing spinor of the background. 
Taking into account the projections satisfied by the spinor and the normalization condition $H^{{1\over 4}}\,\,\epsilon^{\dagger}\epsilon=1$, we get the following expression for ${\cal K}$:
\beq
{\cal K}\,=\,e^{01}\,\wedge\big(\,e^{23}-e^{56}-e^{47}\,\big)\,\,.
\label{K-es-D3}
\eeq
The form ${\cal K}$ can be rewritten more compactly if one takes into account that
the K\"ahler form of the internal manifold in the frame (\ref{10-D3-frame}) is just:
\beq
J\,=\,e^{23}\,-\,e^{56}\,-\,e^{47}\,\,.
\eeq
Then, if  ${\rm Vol}({\rm Min}_{1,1})$ is the volume form of the Minkowski part of the metric  (${\rm Vol}({\rm Min}_{1,1})=H^{-{1/2}}\,dx^0\wedge dx^1$), one can write the calibration form ${\cal K}$ as:
\beq
{\cal K}\,=\,{\rm Vol}({\rm Min}_{1,1})\,\wedge\,J\,\,.
\eeq
One can verify that the supersymmetry preserving conditions (\ref{BPS-D3}) can be rewritten as:
\beq
d\,{\cal K}\,=\,{}^*\,d\,{\cal C}_4\,\,,\qquad\qquad
d\Big(\,H^{-{1\over 2}}\,{}^*\,{\cal K}\,\Big)\,=\,0\,\,,
\label{Cal-con-D3}
\eeq
where the star ${}^*$ denotes the Hodge dual in the ten-dimensional geometry (\ref{D3-metric}). 
Notice that the first of the conditions (\ref{Cal-con-D3}) implies the following relation between the RR five-form field strength and the calibrating form:
\beq
F_5\,=\,d\,{\cal K}\,+\,{}^*\,d\,{\cal K}\,\,.
\eeq

\subsection{Integration of the BPS system}

The brane setup studied in this section can be realized in the context of five-dimensional gauged supergravity \cite{Maldacena:2000mw}. In this approach one formulates an ansatz for the 5d metric, scalar and gauge fields in which the different functions depend on one radial variable and, therefore, the corresponding BPS equations are ordinary differential equations. This BPS system was obtained in section 3.2 of \cite{Maldacena:2000mw} and will not be repeated here. After uplifting the ansatz of \cite{Maldacena:2000mw} to ten dimensions, and by performing a suitable change of variables,  one can define two radial coordinates $\rho$ and $\sigma$ in such a way that  a metric of the type (\ref{D3-metric}) and a five-form as in (\ref{F5-total})-(\ref{C4}) are obtained (see refs. \cite{Arean:2008az,Arean:2009gc} for similar analysis in other brane setups).

Although we have not been able to analytically  integrate the BPS system of \cite{Maldacena:2000mw}, it turns out that the different first-order equations can be combined to produce a single second-order differential equation, whose solutions allow one to solve our BPS system (\ref{BPS-D3}) in implicit form. In order to present this solution, let us introduce an auxiliary  function $\tau=\tau(z)$, defined as the solution of the 
following ordinary second-order differential equation:
\beq
{d^2\tau \over d z^2}\,=\,\Big[\,2 e^{2\tau}\,-\,{1\over 2 z}\,\Big]\,{d\tau \over d z}\,\,.
\label{master}
\eeq
Then, one can show that the PDE  (\ref{PD3-D3unflavored}) can be solved in implicit form by taking  $z(\rho,\sigma)$ as the solution of the equation:
\beq
{\rho^2\over e^{2\tau(z)}}\,+\,{\sigma^2\over \sqrt{z}\,{d\tau\over dz}}\,=\,{1\over m^2}\,\,.
\label{implicitD3}
\eeq
The other functions of our ansatz can also be obtained. Indeed, the warp factor $H(\rho, \sigma)$ corresponding to the function $z(\rho,\sigma)$ of (\ref{implicitD3}) is given by:
\beq
H\,=\,{\sqrt{z}\over m^2\,\Big[\,\rho^2\,\sqrt{z}\,e^{-2\tau}\,\Big({d\tau\over dz}\Big)^2\,+\,
\sigma^2\,e^{2\tau}\,\Big]}\,\,,
\label{H-D3}
\eeq
while  the functions $g_1(\rho,\sigma)$ and $g_2(\rho, \sigma)$ are:
\beq
g_1\,=\,{\sigma^4\,e^{2\tau}\over 8\, z}\,\,\Big({d\tau\over d z}\Big)^{-1}\,H\,\,,
\qquad\qquad
g_2\,=\,{\sigma^2\over 8\, m^2\,\sqrt{z}}\,\,\Big({d\tau\over d z}\Big)^{-1}\,\,.
\label{gs-D3}
\eeq
One can verify that $z$, $H$, $g_1$ and $g_2$, as given in eqs. (\ref{master})-(\ref{gs-D3}), satisfy the system of PDEs written in (\ref{BPS-D3}). It is also easy to check that, when $m$ is given by (\ref{m-D3}), the functions (\ref{gs-D3}) give rise to the correct quantization of the flux of $F_5$  for a stack of $N_c$ D3-branes. 
The auxiliary equation (\ref{master}) can be easily integrated numerically. This numerical solution can be used in (\ref{implicitD3})-(\ref{gs-D3}) to get the different functions of our ansatz in the $(\rho,\sigma)$ plane. Before performing this numerical analysis, let us present in the next subsection some approximate results for the region of the background in which the value of the function $\tau$ is large which, as we shall argue, will turn out to correspond to the UV region of the corresponding dual field theory.

\subsubsection{Approximate UV solution}
\label{UV-D3}
For large $\tau$ the master equation (\ref{master}) can be written as:
\beq
{d^2\tau \over d z^2}\,\approx\,\,2 e^{2\tau}\,{d\tau \over d z}\,=\,
{d\over dz}\,\big(\,e^{2\tau}\,\big)\,\,,
\eeq
which can be immediately integrated, namely:
\beq
{d\tau\over d z}\,=\,e^{2\tau}\,+\,c\,\,,
\eeq
with $c$ being a constant of integration. An additional integration gives $\tau$ as a function of $z$:
\beq
e^{2\tau}\,=\,{c\over e^{{2c(z_*-z)}}-1}\,\,,
\label{tau(z)-UV}
\eeq
where $z_*$ is a new constant of integration which represents the value of $z$ for which $\tau\to \infty$. 
By making use of (\ref{tau(z)-UV}) in (\ref{implicitD3}), we arrive at the following implicit relation for $z(\rho,\sigma)$ in the large $\tau$ region:
\beq
\rho^2\,+\,{\sigma^2\over \sqrt{z}\,\,e^{{2c(z_*-z)}}}\,=\,{c\over m^2
\Big[\,e^{{2c(z_*-z)}}-1\,\Big]}\,\,.
\label{implicit-UV}
\eeq
Notice that for large $\tau$ one has $z\approx z_*$ and, according to (\ref{implicit-UV}), we are far from the origin in the $(\rho, \sigma)$ plane. Thus, large $\tau$ corresponds to the UV region of the background, as stated above. Furthermore,  in this deep UV region we can take  $e^{{2c(z_*-z)}}-1\approx 2c(z_*-z)$ on the right hand side of (\ref{implicit-UV})  and, therefore   $z(\rho,\sigma)$  can be approximately written as:
\beq
z\,\approx\,z_*\,-\,{\sqrt{z_*}\over 2m^2}\,\,{1\over \sigma^2\,+\,\sqrt{z_*}\,\,\rho^2}\,\,.
\label{z-UVapprox-D3}
\eeq
Moreover, it follows from (\ref{implicitD3}) that in the deep UV where, at leading order,  $z\approx z_*$ and ${d\tau\over dz}\approx e^{2\tau}$, one has:
\beq
\sigma^2\,+\,\sqrt{z_*}\,\rho^2\,\approx\,{\sqrt{z_*}\over m^2}\,e^{2\tau}\,\,.
\eeq
On the other hand, using these UV estimates in (\ref{H-D3}) and (\ref{gs-D3}), one gets:
\beq
H\approx {z_*\over m^4}\,\,{1\over (\,\sigma^2\,+\,\sqrt{z_*}\,\,\rho^2\, )^2}\,\,,\qquad
g_1\,\approx\,{1\over 8 m^4}\,\,{\sigma^4\over (\sigma^2\,+\,\sqrt{z_*}\,\rho^2\,)^2}\,\,,
\qquad
g_2\,\approx\,{1\over 8 m^4}\,\,{\sigma^2\over \sigma^2\,+\,\sqrt{z_*}\,\rho^2}\,\,.
\eeq
The above results suggest that, in the UV, the different quantities depend on 
 $\sigma$ and $\rho$  through the combination $\sigma^2\,+\,\sqrt{z_*}\,\,\rho^2$. Accordingly, let us introduce the following new variables $u$ and $\alpha$:
\beq
\sigma\,=\,u\,\sin\alpha\,\,,\qquad\qquad
(z_*)^{{1\over 4}}\,\rho\,=\,u\,\cos\alpha\,\,,
\label{u-alpha}
\eeq
with $0\le u<\infty$ and $0\le\alpha \le\pi/2$. 
Since the combination $\sigma^2\,+\,\sqrt{z_*}\,\,\rho^2$ is just $u^2$, we get
that:
\beq
H\approx {z_*\over m^4}\,{1\over u^4}\,\,,
\qquad\qquad
z-z_*\,\approx\,-{\sqrt{z_*}\over 2 m^2\,u^2}\,\,.
\eeq
Using  these results for $H$ and $z$, one immediately verifies that the UV metric becomes:
\bear
&&ds^2_{UV}\,\approx\,{m^2\over \sqrt{z_*}}\,u^2\,
\Big[\,dx^2_{1,1}\,+\,{z_*\over m^2}\,\big(\,(d\theta)^2\,+\,\sin^2\theta\,(d\phi)^2\,
\big)\,\Big]
\,+\,{1\over m^2}\,{(du)^2\over u^2}\,+\rc\rc
&&
+{1\over m^2}\,\Big[\,(d\alpha)^2\,+\,{\sin^2\alpha\over 4}\,
\Big(\,(w^1)^2\,+\,(w^2)^2\,+\,(w^3+\cos\theta\,d\phi)^2\,\Big)\,
+\,\cos^2\alpha\,(d\lambda)^2
\,\Big]\,\,.\qquad
\eear
Notice that the first line corresponds to the metric of an $AdS_5$ space, with two of its directions compactified on an $S^2$. The second line is the metric of a five-sphere fibered over the $S^2$.

\subsection{Probe analysis}
\label{probe-D3}

In order to make contact with the gauge theory dual to the background just studied, let us consider a probe color D3-brane. This probe brane is extended along $(x^0, x^1, \theta, \phi)$ at fixed values of the other coordinates. We will first assume that the worldvolume gauge fields of the D3-brane are not excited. By explicitly computing the DBI and WZ terms for this configuration by using  the metric (\ref{D3-metric}) and the RR four-form potential corresponding to the $F_5$ of (\ref{F5-total})-(\ref{C4}) ($F_5=dC_4$), one realizes that the D3-brane is only at equilibrium if the coordinate $\sigma$ vanishes. Moreover, one can check that $\sigma=0$ is the SUSY locus of the color D3-branes. The calculations leading to these conclusions are the same as those performed in  section 4.1 of \cite{Arean:2008az} for the case with eight supersymmetries and, therefore, they will not be repeated here. 

Let us next consider a color brane located at the no-force point $\sigma=0$ and let us assume that we switch on the worldvolume gauge field $F$ in such a way that its only non-vanishing components are those along the unwrapped directions $x^{\mu}$. By expanding the DBI action in powers of $F$ and looking at the coefficient of the $F^2$ term, we get the holographic expression of the Yang-Mills coupling $g_{YM}$. Again, the calculation is the same as in \cite{Arean:2008az} and gives the result:
\beq
{1\over g_{YM}^2}\,=\,{z(\rho,\sigma=0)\over g_s\,m^2}\,\,.
\label{YM-D3}
\eeq
Notice that the right-hand side of (\ref{YM-D3}) is proportional to the size of the cycle, which is controlled by the function $z$. Moreover, since $z(\rho,\sigma)$ in (\ref{YM-D3}) is evaluated at $\sigma=0$ (\ie\ at the SUSY locus), the remaining radial coordinate $\rho$ plays the role of a holographic coordinate. In the next subsection we will present numerical results for $z(\rho,\sigma)$. It is however possible to extract relevant physical information by using the analytical results of subsection \ref{UV-D3} in the UV region of the geometry. Indeed, by using (\ref{z-UVapprox-D3}) for $\sigma=0$  and large $\rho$ we get:
\beq
{1\over g_{YM}^2}\approx {1\over g_{*}^2}\,-\,{1\over 2 m^4\,g_s\,\rho^2}\,\,,
\label{YM-D3-UV}
\eeq
with $ g_{*}^2= m^2 g_s/z_*$. It follows from (\ref{YM-D3-UV}) that  $g_{YM}$ approaches the fixed-point value $g_{YM}=g_{*}$ in the UV and deviates from it by means of a power law in the holographic coordinate $\rho$. Notice that the running  with a $\rho^{-2}$  power of the right-hand side of  (\ref{YM-D3-UV}) was to be expected from dimensional arguments. It is just the behavior obtained in one-loop perturbation theory if $\rho$ is taken to be proportional to the renormalization energy scale $\mu$. The minus sign of the running term in (\ref{YM-D3-UV}) is also expected: it just means that $g_{YM}$ decreases when we move towards the UV. However, the coefficient of this term does not coincide with the one found in perturbation theory if the naive identification $\rho=2\pi\alpha'\mu$ is used. This fact is not surprising, given  the amount of SUSY preserved by our background.

\subsection{Numerical results}
\label{numerical-unflavored-D3}

\begin{figure}
\centering
\includegraphics[scale=0.7]{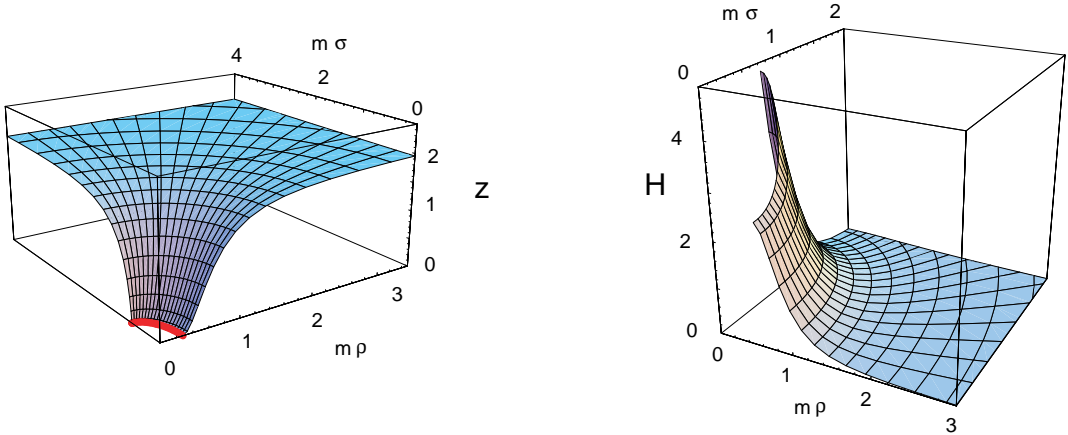}
\caption{On the left we plot $z$, obtained from (\ref{master}) and (\ref{implicitD3}), as a function of the dimensionless variables $m\rho$ and $m\sigma$. We have fixed the UV constants $z_*$ and $c$ of (\ref{implicit-UV}) to the values $z_*=2$ and $c=-1/2$. On the right we represent the warp factor $H$ for these same values. We only show the region where $H$ is monotonic.}
\label{zHunflavoredD3}
\end{figure}

The gauged supergravity solution (\ref{master})-(\ref{gs-D3}) of the BPS system (\ref{BPS-D3}) can be easily evaluated numerically. One must first solve the master equation (\ref{master}) and use the function $\tau(z)$ obtained in this way  to find $z(\rho,\sigma)$ by solving the implicit equation (\ref{implicitD3}). The warp factor $H$ and the functions $g_1$ and $g_2$ are then straightforwardly obtained from (\ref{H-D3}) and (\ref{gs-D3}). 

The result of this numerical analysis for $z(\rho,\sigma)$ and $H(\rho,\sigma)$ has been plotted in figure \ref{zHunflavoredD3}. We notice from this figure that $z(\rho,\sigma)$ grows monotonically when we move away from the origin of the $(\rho,\sigma)$ plane and reaches a constant asymptotic value when $\rho^2+\sigma^2\to \infty$. On the contrary $H(\rho,\sigma)$ increases as $\rho^2+\sigma^2$ decreases until it reaches a maximum and then starts to decrease. This non-monotonic behavior of the warp factor is opposite to the one expected for a good holographic dual of a gauge theory and it indicates to  us that the supergravity solution is not trustable in this IR region. Notice that this problem in the IR is shared by other backgrounds constructed from branes wrapping cycles.

\subsection{Addition of flavor}
\label{flavor-D3}
In this section we will modify the previous setup to include the effect of degrees of freedom corresponding to fields in the fundamental representation of the gauge group (flavors). This can be achieved by including in the setup a new set of D3-branes which introduce a new open string sector. These flavor D3-branes should be extended along the two Minkowski directions, as well as along a two-dimensional non-compact submanifold of the normal bundle $N_4$. Moreover, the flavor branes should be located at a fixed point of the two-cycle ${\cal C}_2$ and of the transverse $\mathbb{R}^2$. The corresponding brane array is:
\begin{center}
\begin{tabular}{|c|c|c|c|c|c|c|c|c|c|c|}
\multicolumn{3}{c}{ }&
\multicolumn{6}{c}
{$\overbrace{\phantom{\qquad\qquad\qquad\qquad\qquad}}^{\text{CY}_3}$}\\
\hline
&\multicolumn{2}{|c|}{$\mathbb{R}^{1,1}$}
&\multicolumn{2}{|c|}{$S^2$}
&\multicolumn{4}{|c|}{$N_4$}
&\multicolumn{2}{|c|}{$\mathbb{R}^{2}$}\\
\hline
$N_c$ D$3$ &$-$&$-$&$\bigcirc$&$\bigcirc$&$\cdot$&$\cdot$
&$\cdot$&$\cdot$&$\cdot$&$\cdot$\\
\hline
$N_f$ D$3$ &$-$&$-$&$\cdot$&$\cdot$&$-$&$-$
&$\cdot$&$\cdot$&$\cdot$&$\cdot$\\
\hline
\end{tabular}
\end{center}
In order to determine the particular embedding of the flavor D3-branes in the normal bundle $N_4$ we will require that the configuration described by the array written above preserves the same four supersymmetries as the unflavored background. In section \ref{micro-D3} we will introduce a general family of such embeddings, which can be briefly described as two-planes embedded in $N_4$ (which has the structure of $\mathbb{R}^4$ fibered over the two-cycle ${\cal C}_2$).  Notice also that the fixed value $\rho_Q$ of the $\rho$ coordinate of the flavor branes represents the distance between the two sets of branes in the transverse $\mathbb{R}^2$, which should be related to the mass $m_Q$ of the matter fields as $m_Q\sim \rho_Q/\alpha'$. 

When $N_f<<N_c$ the flavor branes can be treated as probes which do not alter the unflavored supergravity background. This is the so-called quenched approximation. Here we will concentrate in the limit in which  $N_f$ is large and of the same order as $N_c$ and, therefore,  the effects of the backreaction must be taken into account. Indeed, when $N_f\sim N_c$ one must deal with a coupled gravity plus branes system in which the branes are dynamical objects that act as sources for the different supergravity fields. In particular, the WZ term of the D3-brane worldvolume action couples to the RR four-form potential $C_4$ in the form:
\beq
S_{WZ}\,=\,T_3\,\sum_{N_f}\,\int_{{\cal M}_4}\,\hat C_4\,\,,
\label{D3-WZ}
\eeq
where the hat over $C_4$ denotes its pullback to the D3-brane worldvolume ${\cal M}_4$. In general a term like (\ref{D3-WZ}) is a source for the RR five-form $F_5$ which gives rise to a violation of its Bianchi identity on the flavor  brane worldvolume. Actually, solving the corresponding equations of motion of the coupled system when the flavor branes are  embedded along a given fixed submanifold  ${\cal M}_4$ is a formidable task which, in practice, is not possible to tackle. For this reason we will follow the approach of ref. \cite{CNP} (see \cite{Nunez:2010sf} for a review) and we shall distribute the $N_f\to\infty$ flavor branes along a continuous set of mutually supersymmetric embeddings. In this approach one performs the following substitution in the WZ term:
\beq
\sum_{N_f}\,\int_{{\cal M}_4}\,\hat C_4\,\rightarrow\,\int_{{\cal M}_{10}}\,
\Omega\wedge C_4\,\,,
\eeq
where $\Omega$ is a six-form (the smearing form) which encodes the RR charge density associated to the continuous distribution of flavor branes. Actually, $\Omega$ is just the source in the modified Bianchi identity of $F_5$, namely:
\beq
dF_5\,=\,2\kappa_{10}^2\,T_3\,\Omega\,\,,
\label{D3-Bianchi}
\eeq
where $2\kappa_{10}^2\,=\,(2\pi)^7\,g_s^2\,(\alpha')^4$. It is clear from (\ref{D3-Bianchi}) that, in order to include the backreaction effects, we must modify our ansatz (\ref{F5-total})-(\ref{C4}) for $F_5$. Actually, instead of (\ref{F5-dC4}) we shall take ${\cal F}_5$ as given by:
\beq
{\cal F}_5\,=\,d\,{\cal C}_4\,+\,f_5\,\,,
\eeq
where ${\cal C}_4$ is still given by the ansatz (\ref{C4}) and $f_5$ is a five-form which incorporates the modification of the Bianchi identity (\ref{D3-Bianchi}), namely:
\beq
df_5\,=\,2\kappa_{10}^2\,T_3\,\Omega\,\,.
\label{df5-Omega}
\eeq
As our flavor branes will be located at a fixed value of $\rho=\rho_Q$, it follows that the smearing form $\Omega$ should contain a $\delta(\rho-\rho_Q)$ in its expression. Moreover, it is clear from (\ref{df5-Omega}) that $\Omega$ is a closed form, \ie\ $d\Omega=0$. These two conditions are satisfied if $\Omega$ is of the form:
\beq
2\kappa_{10}^2\,T_3\,\Omega\,=\,\delta(\rho-\rho_Q)\,d\rho\,\wedge d\Lambda\,\,,
\label{Omega-Lambda-D3}
\eeq
where $\Lambda$ is a four-form depending on $\sigma$ and on the angular coordinates. 
Eq. (\ref{Omega-Lambda-D3})  implies that $f_5$ can be taken as:
\beq
f_5\,=\,\Theta(\rho-\rho_Q)\,d\Lambda\,\,,
\eeq
and, therefore,  the total ${\cal F}_5$ is given by:
\beq
{\cal F}_5\,=\,d\,{\cal C}_4\,+\,\Theta(\rho-\rho_Q)\,d\Lambda\,\,.
\label{calF5-flavoredD3}
\eeq
The precise form of $\Lambda$ (and therefore of $\Omega$) is  obtained by computing  the charge density that results after averaging over a particular family of equivalent embeddings that are mutually supersymmetric.  Instead of following this ``microscopic" procedure to determine $\Lambda$, we will follow the ``macroscopic"  approach of \cite{Gaillard:2008wt, Arean:2009gc} and we shall  try to get the form of $\Lambda$ by requiring the preservation of supersymmetry and the  compatibility of the charge density with the metric ansatz (\ref{D3-metric}). Later on, in subsection \ref{micro-D3}, we will precisely characterize the particular set of embeddings that give rise to the charge density determined in the macroscopic formalism.

In the flavored case we shall adopt the same ansatz (\ref{C4}) for the RR potential ${\cal C}_4$ as in the unflavored background. Moreover,  from the expression (\ref{calF5-flavoredD3}) of ${\cal F}_5$ the similarity between ${\cal C}_4$ and $\Lambda$ is quite obvious. For this reason it is quite natural to adopt an ansatz for $\Lambda$ with the same structure as in  (\ref{C4}), namely:
\beq
\Lambda\,=\,L_1(\sigma)\,w^1\wedge w^2\wedge (w^3+\cos\theta d\phi)\,\wedge d\lambda\,+\,L_2(\sigma)\,\Omega_2\wedge (w^3+\cos\theta d\phi)\wedge d\lambda\,\,,
\eeq
where $L_1$ and $L_2$ are functions of the coordinate $\sigma$ to be determined. Since $d\Lambda$ is given by:
\beq
d\Lambda\,=\,\big(L_2-L_1\big)\,\Omega_2\wedge w^1\wedge w^2\wedge d\lambda\,+\,
\Big(\,\dot L_1\,w^1\wedge w^2\,+\,\dot L_2\,\Omega_2\,\Big)\wedge d\sigma\wedge
\big(w^3+\cos\theta d\phi\,\big)\wedge d\lambda\,\,,\qquad
\label{dLambda-D3}
\eeq
it follows  that the flavored BPS ansatz for ${\cal F}_5$ can be obtained from the unflavored one after performing the following substitutions:
\bear
&&\dot g_i\to \dot g_i\,+\,\dot L_i\,\Theta(\rho-\rho_Q)\,\,,\qquad (i=1,2)\,\,,\rc\rc
&&g_2-g_1\to g_2-g_1\,+\,(L_2-L_1)\,\Theta(\rho-\rho_Q)\,\,.
\label{g-L}
\eear
One can now repeat the supersymmetry analysis for the new ansatz. Obviously the flavored BPS system is obtained by performing the substitutions (\ref{g-L}) in (\ref{BPS-D3}), namely: 
\bear
&&z'\,=\,8m^2\,{\sqrt{z}\over \rho\sigma^2}\,\big[\, g_2\,-\,g_1
\,+\,(L_2-L_1)\,\Theta(\rho-\rho_Q)\,\big]\,\,,\rc\rc
&&\dot z\,=\,{m^2\sigma \over \sqrt{z}}\, H\,\,,\rc\rc
&&g'_1\,=\,{1\over 8}\,{\rho\sigma^3\over \sqrt{z}}\,\,\dot H\,-\,{m^2\over 16}\,
{\rho\sigma^4\over z^2}\,\,H^2\,\,,\rc\rc
&&
\dot g_1\,=\,-{1\over 8}\,{\rho\sigma^3\over  z}\,\,H'\,+\,
m^2\,{\sigma H\over z^{{3\over 2}}}\,(\,g_2\,-\,g_1
\,+\,(L_2-L_1)\,\Theta(\rho-\rho_Q)\,)\,-\,\dot L_1\,\Theta(\rho-\rho_Q)\,\,,\rc\rc
&& g_2'\,=\,-{1\over 4}\,{\rho\sigma^2\over \sqrt{z}}\,H\,\,,\rc\rc
&&\dot{ g_2}\,=\,{2\over \sigma}\,(\, g_2\,-\,g_1\,)\,-\,
\big[\,\dot L_2\,-\,{2\over \sigma}\,(L_2-L_1)\,\big]\,\Theta(\rho-\rho_Q)\,\,.
\label{BPS-D3-flavored}
\eear
By analyzing the compatibility of the different equations in the  system (\ref{BPS-D3-flavored}) one discovers that, in general, the $(\rho, \sigma)$ crossed derivatives of the functions $g_1$ and $g_2$ are not equal. Indeed, one can prove that:
\beq
\partial_{\rho}\,\dot g_1\,-\,\partial_{\sigma}\,g_1'\,=\,
\partial_{\sigma}\,g_2'\,-\,\partial_{\rho}\,\dot g_2\,=\,
-{1\over \sigma}\,
\big[\,2(L_2\,-\,L_1)\,-\,\sigma \dot L_2\,\big]\,
\delta(\rho-\rho_Q)\,\,.
\label{crossed}
\eeq
Therefore, in order to get rid of this unwanted singularity  we must require the following condition on $L_1$ and $L_2$:
\beq
\dot L_2\,=\,{2\over \sigma}\,\big(\,L_2\,-\,L_1\,\big)\,\,.
\label{L12-eq-D3}
\eeq
Notice that, after imposing (\ref{L12-eq-D3}), the last equation in (\ref{BPS-D3-flavored}) is the same as the one in the unflavored system (\ref{BPS-D3}). Furthermore, one can verify that any solution of the BPS system (\ref{BPS-D3-flavored})  and the compatibility condition (\ref{L12-eq-D3}) also solves the equations of motion of gravity with source terms coming from the smeared branes.  Moreover, in this case one can also write a single PDE for $z(\rho,\sigma)$, which now has a source term parametrized by $L_2-L_1$, namely:
\beq
2\rho\,z\,\sqrt{z}\,\big(\,\sigma\,\ddot z\,+\,\dot z\,\big)\,=\,\sigma\,\big(\,\rho\,z'^2\,
-\,2z\,z'\,-\,2\rho z\,z''\,\big)\,+\,{16m^2(L_2-L_1) \,z^{{3\over 2}}\over \sigma}\,\,
\delta(\rho-\rho_Q)\,\,.
\label{PD3-D3flavored}
\eeq
The compatibility condition (\ref{L12-eq-D3}) allows one  to determine $L_2(\sigma)$ in terms of $L_1(\sigma)$. In principle, there are and infinite number of solutions to (\ref{L12-eq-D3}). However, it turns out that there is a particularly simple solution which satisfies certain regularity conditions which should be satisfied on general physical grounds. Since the complete analysis is very similar to the one performed in section 3.3 of \cite{Arean:2009gc}, it will not be repeated here. Instead, we will write down the solution and we will verify that it satisfies some of these physical requirements. The solution of (\ref{L12-eq-D3})  we will be interested in from now on is the one obtained by taking $L_1(\sigma)=0$, namely:
\beq
L_1\,=\,0\,\,,\qquad\qquad
L_2\,=\,C\,\sigma^2\,\,,
\label{L12-D3-simple}
\eeq
where $C$ is a constant. By using these values of $L_1$ and $L_2$ in eqs. (\ref{dLambda-D3}) and (\ref{Omega-Lambda-D3}) one can readily get the corresponding expression of the smearing form $\Omega$, \ie:
\beq
\Omega\,=\,{C\over 2 \kappa_{10}^2\,T_3}\,\delta(\rho-\rho_Q)\,
d\rho\wedge d\lambda\wedge \Omega_2\,\wedge \Big[\,\sigma^2\,w^1\wedge w^2\,+\,
2\sigma\,d\sigma\wedge w^3\,\Big]\,\,,
\label{Omega-special-D3}
\eeq
which, in terms of the vielbein basis (\ref{10-D3-frame}), takes the following form:
\beq
\Omega\,=\,{4 m^2 C\over 2 \kappa_{10}^2\,T_3}\,
{\delta(\rho-\rho_Q)\over \rho\,\sqrt{z\,H}}\,\,e^{23}\,\wedge\big[\,
e^{56}\,+\,e^{47}\,\big]\,\wedge e^{89}\,\,.
\label{Omega-es-D3}
\eeq
An important test that the smearing form $\Omega$ must pass is the regularity and positivity of the mass density distribution of the flavor brane sources. This mass distribution density can be obtained by looking at the smeared version of the DBI action of the flavor branes. Indeed, taking into account that, for a SUSY configuration, the induced volume form is just the pullback of the calibration form ${\cal K}$ and that the smearing is performed by taking the wedge product with $\Omega$, this DBI action takes the form:
\beq
S^{flavor}_{DBI}\,=\,-T_3\,\int_{{\cal M}_{10}}\,\Omega\wedge {\cal K}\,\,.
\eeq
Thus,  $\Omega\wedge {\cal K}$ can be interpreted as the mass distribution of the system of flavor branes which, being a ten-form in a ten-dimensional space,  is proportional to the volume form  ${\rm Vol}\,({\cal M}_{10})$ of the ten-dimensional manifold. Actually,  from (\ref{Omega-es-D3}) and  the expression of ${\cal K}$ in (\ref{K-es-D3}) one straightforwardly gets:
\beq
\Omega\wedge {\cal K}\,=\,-{4 m^2\,C\over \kappa_{10}^2\,T_3}\,\,
{\delta(\rho-\rho_Q)\over \rho\sqrt{z}\,H}\,\,{\rm Vol} \big (\,{\cal M}_{10}\,\big)\,\,.
\label{Omega-wedge-K-D3}
\eeq
The function multiplying ${\rm Vol}\,({\cal M}_{10})$ in (\ref{Omega-wedge-K-D3}) represents the mass density of the $\rho=\rho_Q$ slice of the space in which we  are smearing the flavor branes.  The positivity of this mass density implies that the constant $C$ must be negative. In the next subsection we will find a family of supersymmetric embeddings which gives rise to the charge distribution (\ref{Omega-special-D3}) and we will identify the constant $C$ with the parameters of this set of smeared branes.

\subsubsection{A microscopic interpretation}
\label{micro-D3}
In order to characterize a family of embeddings of  the flavor D3-branes that preserve the supersymmetries of the background,  let us parametrize the one-forms $w^i$ in terms of three angles $\hat\theta$, $\hat\phi$ and $\hat\psi$ as follows:
\bear
w^1 & = & \cos\hj\,\td\hth+\sin\hj\,\sin\hth\,\td\hf\,\,, \rc
w^2 & = & \sin\hj\,\td\hth-\cos\hj\,\sin\hth\,\td\hf\,\,, \rc
w^3 & = & \td\hj+\cos\hth\,\td\hf\,\,,
\label{w123}
\eear
with $0\le \hat\theta\le \pi$, $0\le\hat\phi<2\pi$, $0\le\hat\psi \le 4\pi$.  Let us next introduce the coordinates $y^i$ $(i=1,\cdots, 4)$ as:
\beq
y^2+iy^1\,=\,\sigma\,\cos\Big({\hat \theta\over 2}\Big)\,
e^{{i\over 2}\,\big(\hat\psi+\hat\phi\big)}\,\,,\qquad
y^4+iy^3\,=\,\sigma\,\sin\Big({\hat \theta\over 2}\Big)\,
e^{{i\over 2}\,\big(\hat\psi-\hat\phi\big)}\,\,.
\label{y-def}
\eeq
One can verify by direct calculation that:
\beq
\sum_i\,\big(\,dy^i\,\big)^2\,=\,(d\sigma)^2\,+\,{\sigma^2\over 4}\,\,
\sum_i\,\big(\,w^i\,\big)^2\,\,,
\eeq
which shows that the normal bundle $N_4$ has the structure of $\mathbb{R}^4$ fibered over the $(\theta,\phi)$ two-sphere and that the $y^i$'s are just the cartesian coordinates of this four-plane. Moreover, one can also check that:
\beq
dy^1\wedge dy^2\,+\,dy^3\wedge dy^4\,=\,-{1\over 4}\,\,
\Big[\,\sigma^2\,w^1\wedge w^2\,+\,2\sigma\,d\sigma\,\wedge w^3\,\Big]\,\,.
\label{d2ys}
\eeq
Using (\ref{d2ys}) in (\ref{Omega-special-D3}) one immediately proves that the smearing form $\Omega$ can be written in terms of the coordinates $y^i$ as:
\beq
\Omega\,=\,-{2C\over \kappa_{10}^2\,T_3}\,\,\delta(\rho-\rho_Q)\,
d\rho\wedge d\lambda\wedge \Omega_2\wedge
\big[\,dy^1\wedge dy^2\,+\,dy^3\wedge dy^4\,\big]\,\,.
\label{Omega-ys-D3}
\eeq
The standard way to determine if a given D-brane embedding is supersymmetric is by verifying the fulfillment of the kappa symmetry condition on the Killing spinors of the background. Equivalently, one can demonstrate the supersymmetric nature of the embedding by checking that the pullback of the generalized calibration form on the D-brane worldvolume ${\cal M}_4$ is equal to the induced volume element. In our case this calibration condition reads:
\beq
\hat \cK=\textrm{Vol}({\cal M}_4)\,\,,
\label{eqn:kappa-sym}
\eeq
where ${\cal K}$ is the four-form written in (\ref{K-es-D3}). According to the brane array in the beginning of subsection \ref{flavor-D3}, let us consider a D3-brane which sits at a particular  point of the $(\theta, \phi)$ two-sphere and that is localized in the $\mathbb{R}^2$ plane parametrized by $(\rho,\lambda)$. In addition, the D3-brane is extended along a codimension two surface of the four-dimensional plane spanned by the $y^i$ coordinates. The pullback of  ${\cK}$ for these embeddings  takes the form:
\beq
\hat \cK={1 \over \sqrt{z}}
\td x^0 \wedge \td x^1 \wedge \left[\td y^1 \wedge \td y^2 + \td y^3 \wedge \td y^4 \right]
\,\,,
\eeq
where the pullback of the $y^i$ coordinates is understood. 
Moreover, let us assume that the embedding in the normal bundle $N_4$ is described by linear relations of the type:
\beq
y^3\,=\,a_1\,y^1\,+\,b_1\,\,,\qquad
y^4\,=\,a_2\,y^2\,+\,b_2\,\,,
\eeq
which represent a two-plane in $\mathbb{R}^4$.  For these linear embeddings one has:
\beq
\hat \cK={1 \over \sqrt{z}}\,\,(\,1+a_1\,a_2\,)\,
\td x^0 \wedge \td x^1 \wedge \td y^1 \wedge \td y^2\,\,.
\eeq
On the other hand, the induced metric  on the D3-brane worldvolume can be written as:
\begin{equation}
\td\hat{s}^2=H^{-1/2}\td x_{1,1}^2+ {H^{1/2} \over \sqrt{z}} \left[ \left(1+ a_1^2 \right)(\td y^1)^2 +
 \left(1+ a_2^2 \right)(\td y^2)^2 \right] \, ,
\end{equation}
and the corresponding volume form is
\begin{equation}
\textrm{Vol}({\cal M}_4)={1 \over \sqrt{z}} \, \sqrt{\left(\,1+ a_1^2\, \right)\left(\,1+ a_2^2\, \right)} \, 
\td x^0 \wedge \td x^1 \wedge \td y^1 \wedge \td y^2 \, \,.
\end{equation}
It is now straightforward to prove that  the calibration condition holds if $a_1=a_2$. Let us parametrize:
\beq
a_1=a_2\,=\,-\cot \gamma\,\,,
\eeq
where $\gamma$ is a constant. Then, the embedding in the $y^i$ hyperplane  is characterized by the equations:
\bear
&&f_1\,\equiv\,\cos\gamma\,y^1\,+\,\sin\gamma\,y^3\,-\,c_1\,=\,0\,\,,\rc\rc
&&f_2\,\equiv\,\cos\gamma\,y^2\,+\,\sin\gamma\,y^4\,-\,c_2\,=\,0\,\,,
\label{planes-D3}
\eear
where $c_1$ and $c_2$ are new constants. Eq. (\ref{planes-D3}) defines a family of embeddings parametrized by three parameters ($\gamma$, $c_1$ and $c_2$).\footnote{Actually, there is a much larger family of calibrated embeddings for this background. If we complexify  the $y^{i}$ coordinates as $z^{1,2}=y^{2,4}+iy^{1,3}$, then any submanifold in $N_4$ defined by a holomorphic relation of the type $z^{2}=f(z^1)$ satisfies (\ref{eqn:kappa-sym}). In particular, a linear relation  such as $\alpha z^1\,+\,\beta z^2\,=\,{\rm constant}$ defines a complex line in $\mathbb{C}^2$ which generalizes (\ref{planes-D3}). However, this more general family of planes gives rise to the same $\Omega$ as the one obtained from (\ref{planes-D3}) and, thus,  we will not consider any of these more general embeddings. } 
Notice that changing  $\gamma$ by $\gamma+\pi$ is equivalent to taking $c_i\to -c_i$ in (\ref{planes-D3}). Thus, we will  take $\gamma$ in the interval $0\le\gamma <\pi$.

Let us now compute the charge density six-form $\Omega$ that results after averaging over this family of embeddings. We will homogeneously distribute the D3-brane embeddings over the $(\theta,\phi)$ two-sphere and we will consider branes with a given value $\rho_Q$ of the radial coordinate of the transverse $\mathbb{R}^2$ and distributed over the angular direction parametrized by $\lambda$. Thus, 
 $\Omega$ can be written as:
\beq
\Omega\,=\,\delta(\rho-\rho_Q)\,d\rho\,\wedge\,{d\lambda\over 2\pi}\,\wedge\,
{\sin\theta d\theta\wedge d\phi\over 4\pi}\,\wedge \Gamma\,\,,
\label{Omega-microD3}
\eeq
where we have normalized the distributions in  the coordinates $\lambda$ and  $(\theta,\phi)$ by dividing by the volume of $S^1$ and $S^2$ respectively. In (\ref{Omega-microD3}) $\Gamma$ is the two-form  density obtained after averaging over the family of planes (\ref{planes-D3}). It can be written as the integral of the volume element of the complement of the two-planes (\ref{planes-D3}) in $N_4$, namely \cite{Bigazzi:2008zt}:
\beq
\Gamma\,=\,\int d\mu\,\delta(f_1)\,\delta(f_2)\,\big[\,df_1\wedge df_2\,\big]\,\,,
\label{Gamma-D3}
\eeq
where $d\mu$ is an integration measure depending on the parameters $\gamma$, $c_1$ and $c_2$ and the  exterior derivative only acts on the $y^i$ variables of the $f_j$. The natural integration measure $d\mu$ for the family of planes (\ref{planes-D3}) is:
\beq
d\mu\,=\,n_f\,{d\gamma\over \pi}\,dc_1\,dc_2\,\,,
\eeq
with $n_f$ being the density of flavor branes.  The integrals over $c_1$ and $c_2$ in (\ref{Gamma-D3}) can be immediately performed by making use of the two delta functions . Moreover, since:
\beq
df_1\wedge df_2\,=\,\cos^2\gamma\,dy^1\wedge dy^2\,+\,
\sin^2\gamma\,dy^3\wedge dy^4\,+\,
\sin\gamma\cos\gamma\,\big[\,dy^1\wedge dy^4\,-\,dy^2\wedge dy^3\,\big]\,\,,
\eeq
we get:
\beq
\int_0^\pi\,d\gamma\,\big[\,df_1\wedge df_2\,\big]\,=\,{\pi\over 2}\,
\big[\,dy^1\wedge dy^2\,+\,dy^3\wedge dy^4\,\big]\,\,.
\eeq
Thus, the two-form $\Gamma$ can be written as:
\beq
\Gamma\,=\,{n_f\over 2}\,\,
\big[\,dy^1\wedge dy^2\,+\,dy^3\wedge dy^4\,\big]\,\,,
\label{Gamma-ys}
\eeq
and the resulting $\Omega$ is, indeed, of the form (\ref{Omega-ys-D3}) with the constant $C$ related to the density
$n_f$ as:
\beq
C\,=\,-2\kappa_{10}^2\,T_3\,\,{n_f\over 64\pi^2}\,\,.
\label{C-n_f-D3}
\eeq
\subsubsection{Numerical results}

\begin{figure}
\centering
\includegraphics[scale=0.7]{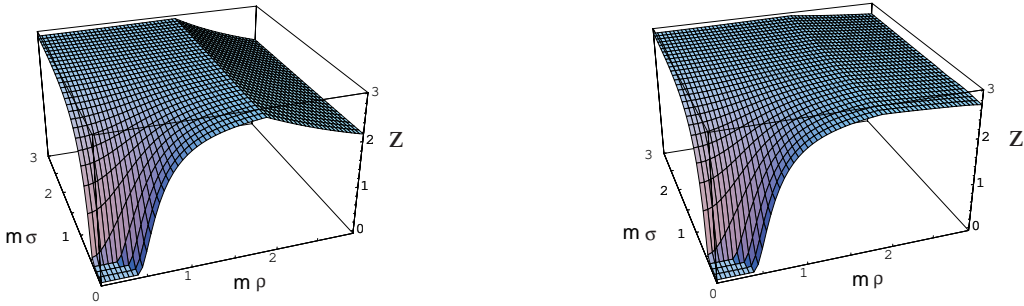}
\caption{Plots of $z(\rho,\sigma)$ for the flavored background of D3-branes. The flavor branes are located at $m\,\rho_Q=1.8$ for two different values of $x\equiv{\pi n_f\over m^2N_c}$. The plot on the left (right) corresponds to $x=2$ ($x=1/2$). The solution for $\rho<\rho_Q$ is the unflavored one  with UV parameters $z_*=3$ and $c=-1.15$. 
}
\label{zflavoredD3}
\end{figure}

Let us now present the results obtained by the numerical integration of the BPS system 
(\ref{BPS-D3-flavored}). As in \cite{Angel,Arean:2008az,Ramallo:2008ew,Arean:2009gc}, our strategy will be to integrate the PDE equation (\ref{PD3-D3flavored}) for $z(\rho,\sigma)$. Actually, after using the values of $L_1$ and $L_2$ written in (\ref{L12-D3-simple}) and the relation  between  $C$ and $n_f$ displayed in eq. (\ref{C-n_f-D3}), the PDE (\ref{PD3-D3flavored}) takes the form:
\beq
2\rho\,z\,\sqrt{z}\,\big(\,\sigma\,\ddot z\,+\,\dot z\,\big)\,=\,\sigma\,\big(\,\rho\,z'^2\,
-\,2z\,z'\,-\,2\rho z\,z''\,\big)\,-\,{1\over m^2}\,{\pi\, n_f\over N_c}\,
\sigma\,z^{{3\over 2}}\,\delta(\rho-\rho_Q)\,\,.
\label{PDE-D3flavored-with-constant}
\eeq
Since the system (\ref{BPS-D3-flavored}) reduces to the unflavored one for $\rho<\rho_Q$, we can assume that the solution of (\ref{PDE-D3flavored-with-constant}) is given by 
(\ref{master})-(\ref{implicitD3}) in this region. At $\rho=\rho_Q$ the function $z$ is continuous while, according to the first equation in  (\ref{BPS-D3-flavored}), $z'$ has a
discontinuity  given by:
\beq
z'(\rho_Q+\epsilon, \sigma)\,-\,z'(\rho_Q-\epsilon, \sigma)\,=\,
-{\pi\, n_f\over 2 m^2\, N_c}\,\,{\sqrt{z(\rho_Q, \sigma)}\over \rho_Q}\,\,.
\label{jump-D3}
\eeq
We have integrated (\ref{PDE-D3flavored-with-constant}) by using the continuity at $\rho=\rho_Q$  and the change in derivative (\ref{jump-D3}) as boundary conditions. The results are shown in figure \ref{zflavoredD3} for two values of $n_f$. In our numerical analysis we have been able to integrate $z$ up to a finite value of $\rho$, after which the solution becomes highly oscillatory and unstable. As is evident from figure \ref{zflavoredD3}, the function $z$ has a wedge shape at the position $\rho=\rho_Q$ of the flavor branes. Thus, there will be a curvature singularity at this point, which is actually required by the Einstein equation in order to match a similar term coming from the energy-momentum tensor of the flavor brane source. In general, the addition of flavor makes $z(\rho,\sigma)$ grow slower or to decrease as we move towards the UV for $\rho>\rho_Q$. In view of the relation (\ref{YM-D3}) between $z$ and the YM coupling, this behavior was  to be expected since, contrary to what happens with the non-abelian color fields, the flavor matter fields always make the coupling  grow in the UV. Thus, our flavored geometry correctly encodes this non-trivial property of the renormalization of gauge theories.

\section{ ${\cal N}=2$ 3d SQCD   from wrapped D4-branes}
\label{D4-section}

In this section we will analyze some gravity duals of three-dimensional ${\cal N}=2$ supersymmetric gauge theories which are generated by D4-branes wrapped along two-cycles of a Calabi-Yau threefold. Our brane setup, including flavor,  is very similar to the one studied in section \ref{D3-section}, namely:
\begin{center}
\begin{tabular}{|c|c|c|c|c|c|c|c|c|c|c|}
\multicolumn{3}{c}{ }&
\multicolumn{8}{c}
{$\overbrace{\phantom{\qquad\qquad\qquad\qquad\qquad}}^{\text{CY}_3}$}\\
\hline
&\multicolumn{3}{|c|}{$\mathbb{R}^{1,2}$}
&\multicolumn{2}{|c|}{$S^2$}
&\multicolumn{4}{|c|}{$N_4$}
&\multicolumn{1}{|c|}{$\mathbb{R}$}\\
\hline
$N_c$ D$4$ &$-$&$-$&$-$&$\bigcirc$&$\bigcirc$&$\cdot$
&$\cdot$&$\cdot$&$\cdot$&$\cdot$\\
\hline
$N_f$ D$4$ &$-$&$-$&$-$&$\cdot$&$\cdot$&$-$
&$-$&$\cdot$&$\cdot$&$\cdot$\\
\hline
\end{tabular}
\end{center}

We will be working in the context of  type IIA supergravity and we will adopt the following ansatz for the metric and dilaton in the string frame:
\bear 
&&ds^2_{10}\,=\,H^{-{1\over 2}}\,
\Big[\,dx^{2}_{1,2}\,+\,{z\over m^2}\,\big(\,(d\theta)^2\,+\,\sin^2\theta\, (d\phi)^2\,\big)
\,\Big]\,+\,\rc\rc
&&\qquad
+\,H^{{1\over 2}}\,\Big[\,{1\over z^{{1\over 2}}}\,\Big(\,(d\sigma)^2\,+\,
{\sigma^2\over 4}\,
\big[\,(w^1)^2\,+\,(w^2)^2\,+\, (w^3+\cos\theta\, d\phi)^2\,\big]\,\Big)
+\,
(d\rho)^2\,\Big]\,\,,\qquad\qquad\rc\rc
&&e^{\Phi}\,=\,H^{-{1\over 4}}\,\,,
\label{D4-metric}
\eear
where $-\infty<\rho<+\infty$ and the parameter $m$ is given by:
\beq
{1\over m^2}\,=\, \big(\,8\pi g_sN_c\,\big)^{{2\over 3}}\alpha'\,\,.
\label{m-D4}
\eeq
The similarity of the metric in (\ref{D4-metric}) and the one corresponding to the D3-brane case in (\ref{D3-metric})  is manifest.  However, notice that, in this case, the background should be endowed with an RR four-form field strength $F_4$. In analogy with the D3-brane case, we will take
$F_4$ to be of the form:
\beq
F_4\,=\,d\,C_3\,+\,\Theta(\rho-\rho_Q)\,d\Lambda\,\,,
\label{F4-ansatz}
\eeq
with $C_3$ and $\Lambda$ being given by:
\bear
&&C_3\,=\,g_1\,w^1\wedge w^2 \wedge \left(w^3 + \cos \theta d\phi \right)+g_2\, \Omega_2  \wedge \left(w ^3 + \cos \theta d\phi \right) \, ,\rc\rc
&&\Lambda\,=\,L_1(\sigma)\,w^1\wedge w^2 \wedge \left(w^3 + \cos \theta d \phi \right)\,+\,L_2(\sigma)\,
\,\Omega_2 \wedge \left(w^3 + \cos \theta d \phi \right) \,\,,
\label{C3-Lambda}
\eear
where $\Omega_2$ is the volume form of the $S^2$ written in (\ref{Omega2}) and $g_i(\rho, \sigma)$ and $L_i(\sigma)$ are functions to be determined. Notice that we have already included in (\ref{F4-ansatz}) the term that modifies the Bianchi identity  for $F_4$ due to the presence of the flavor D4-branes at $\rho=\rho_Q$. Indeed, by a simple  calculation from (\ref{F4-ansatz}) we get:
\beq
dF_4\,=\,\delta(\rho-\rho_Q)\,d\rho\,\wedge d\Lambda\,\,. 
\label{Bianchi-D4-ansatz}
\eeq
The source term on the right-hand side of (\ref{Bianchi-D4-ansatz}) comes from the coupling of the D4-brane to the RR fields through the WZ term of the worldvolume action, 
$S_{WZ}=T_4\int \Omega\wedge C_5$, where $\Omega$ is the RR   five-form charge density  of  the flavor brane distribution  and $C_5$ is the five-form potential of $F_6$ (
$F_4={}^*\,F_6$). The corresponding Bianchi identity with source is:
\beq
dF_4\,=\,2\kappa_{10}^2\,T_4\,\Omega\,\,,
\eeq
which allows one to identify $\Omega$ for our ansatz with:
\beq
2\kappa_{10}^2\,T_4\,\Omega\,=\,\delta(\rho-\rho_Q)\,d\rho\,\wedge d\Lambda\,\,.
\eeq
In order to find the system of BPS equations for a type IIA background with metric, dilaton and RR four-form as in our ansatz, let us consider the following vielbein basis for the metric (\ref{D4-metric}):
\begin{eqnarray} 
\label{10d-frame-D4}
&& 
e^{0,1,2}= H^{-\frac{1}{4}} \,dx^{0,1,2} \ ,  \quad  \quad  
e^3= \frac{H^{-\frac{1}{4}} }{m}\,\sqrt{z} \,d \theta \ , \quad  \quad 
e^4 = \frac{H^{-\frac{1}{4}} }{m}\,\sqrt{z}\sin \theta \, d\phi \, ,
\nonumber\\  \nonumber\\
&& 
e^5= \frac{H^{\frac{1}{4}}}{z^{{1\over 4}}} \, d\sigma  \ , \quad  \quad  \quad  \quad  \,\,\,
e^6 =  \frac{H^{\frac{1}{4}}}{2 z^{{1\over 4}}} \, \sigma\, w^1 \ , \quad  \quad \,\,\, 
e^7= \frac{H^{\frac{1}{4}}}{2 z^{{1\over 4}}} \, \,\sigma \, w^2  \,\,,
\\ \nonumber\\
&& 
e^8= \frac{H^{\frac{1}{4}}}{2 z^{{1\over 4}}} \, \,\sigma\,  \left(w^3+\cos \theta \, d\phi\right)
\ , \quad  \quad  \quad \quad  \quad \quad  \quad \,\,\,\,
e^9=H^{\frac{1}{4}} \, d\rho \, .
\nonumber 
\end{eqnarray}
We will now impose the following  projections on the Killing spinors of type IIA supergravity:
\begin{equation} 
\Gamma_{3467}\,\epsilon\,=\,\epsilon \, , \quad \quad  \Gamma_{3458}\,\epsilon\,=\,\epsilon\,\,,   \quad  \quad  \Gamma_{56789}\,\epsilon\,= \, \epsilon \, , 
 \label{D4-projections} 
\end{equation}
where the indices of the constant Dirac matrices refer to the basis (\ref{10d-frame-D4}). 
Notice that the projections (\ref{D4-projections}) leave four supercharges unbroken, corresponding to a three-dimensional theory with ${\cal N}=2$ supersymmetry. 
The corresponding system of  BPS equations is:
\begin{eqnarray} 
&& z'\,=\,8 \, m^2 \, {\sqrt{z} \over \sigma^2} \, 
\Big[g_2-g_1+\left(L_2-L_1\right)\,\Theta (\rho - \rho_Q) \Big]\,, \nonumber \\ \rc
&& \dot z\,= \,{m^2 \, \sigma\over \sqrt{z}}\, H \,, 
\nonumber \\ \rc
&& g_1'\,=\,\frac{1}{8} \, {\sigma^3 \over \sqrt{z}} \, \dot{H}\,-\, \frac{m^2}{16}\,\frac{\sigma^4}{z^2}\, H^2 \, , 
\nonumber \\ \rc
&& \dot{g_1}\, =\, -\, \frac{1}{8}\, \frac{\sigma^3}{z} \, H' \,+ \, m^2 \, \frac{\sigma H}{z^{\frac{3}{2}}}\,
\Big[g_2-g_1+\left(L_2-L_1\right)\,\Theta (\rho - \rho_Q) \Big] -\dot{L}_1\,\Theta (\rho - \rho_Q) \, ,
\nonumber \\ \rc
&&g_2'\,=-\, \frac{1}{4} \, {\sigma^2 \over \sqrt{z}}\, H \,, \nonumber \\ \rc
&& \dot{ g_2}\,=\,{2 \over \sigma}\,(g_2-g_1) -\Big[\,\dot{L}_2\, - {2 \over \sigma} \,\left(L_2-L_1\right)\Big]\,\Theta (\rho - \rho_Q) \,.
 \label{flavor-BPS-D4}
\end{eqnarray}
Combining the equations of \eqref{flavor-BPS-D4} in different ways we can check that the cross derivatives of both $g_1$ and $g_2$ depend on the ordering of the differentiation.  
In order to avoid this unwanted singularity we need to   impose again (\ref{L12-eq-D3}),
which implies that the $L_i$'s drop from the last equation in (\ref{flavor-BPS-D4}) which, therefore,  is the same in the flavored and unflavored cases.  Moreover, any solution of 
\eqref{flavor-BPS-D4} with $L_1$ and $L_2$ satisfying \eqref{L12-eq-D3} also solves the equations of motion of 
the gravity plus (smeared) branes system while, as in the D3-brane case, we can write a single PDE for the function $z(\rho,\sigma)$, which has a source term parametrized by $L_2\,-\,L_1$, namely:
\begin{equation} 
2\, z \, \sqrt{z} \, \left( \sigma \ddot{z} \, + \, \dot{z} \right)\, =\, \sigma \, \left( {z'}^2 -\, 2\, z\, z'' \right) \,+\,
\frac{16 m^2 z^{3/2}}{\sigma} \,\left(L_2 - L_1\right)\, \, \delta(\rho-\rho_Q)  \, .
\label{flavor-PDE-D4}
\end{equation}
As in the D3-brane case, the BPS system (\ref{flavor-BPS-D4}) can be recast in terms  of the appropriate generalized calibration form ${\cal K}$. In this case  ${\cal K}$ is a five-form, defined in terms of bilinears of the Killing spinors as follows:
\begin{equation}
{\cal K}\, = \, \frac{1}{5!}\,\,H^{{1\over 4}}\, e^{a_1 \cdots  a_5} \, \epsilon^{\dagger} \, \Gamma_{11}\,\Gamma_{a_1 \cdots  a_5} \, \epsilon \, ,
\end{equation}
where $\epsilon$ is normalized as $H^{{1\over 4}}\, \epsilon^{\dagger} \, \epsilon=1$. 
After using \eqref{D4-projections} we get the following expression for ${\cal K}$:
\begin{equation}
{\cal K}\,=\, e^{012} \wedge \left(e^{34} \,- \,e^{58}\, -\, e^{67} \right)\, . 
\label{K-D4}
\end{equation}
Using the expression (\ref{K-D4})  for  ${\cal K}$, as well as the ansatz  (\ref{F4-ansatz})-(\ref{C3-Lambda}) for $F_4$, we can easily verify that the following two conditions: 
\begin{equation} 
d\left(H^{-\frac{1}{4}} {}^*\,{\cal K} \right) \,=\,0\,\,, \qquad  \qquad 
F_4\,=\,\, {}^* \, d\left(H^{\frac{1}{4}} \, {\cal K} \right)  \, ,
\label{cal-conditions-D4}
\end{equation}
reduce to the BPS system (\ref{flavor-BPS-D4}). In (\ref{cal-conditions-D4}) the Hodge dual is taken with respect to the string frame metric  in (\ref{D4-metric}).

\subsection{The unflavored solution}
In the unflavored case, the setup we are studying here can be reduced to the one analyzed in \cite{Maldacena:2000mw}, where a configuration of M5-branes wrapping a two-cycle and preserving four supercharges was considered. By compactifying one of the worldvolume directions of the M5-brane and dimensionally reducing along it, the setup of \cite{Maldacena:2000mw} reduces to the one studied here. In \cite{Maldacena:2000mw}  seven-dimensional gauged supergravity was employed to find a system of first-order BPS equations for the functions of the ansatz. These functions depend on a single radial variable and, therefore, the BPS system of \cite{Maldacena:2000mw}  is a system of ordinary differential equations. As in section \ref{D3-section}, one can perform a change of variables  which converts the ansatz of \cite{Maldacena:2000mw} (uplifted to ten-dimensions)  into our two-variable ansatz.  Moreover, this relation can be used to find a solution of the BPS system (\ref{flavor-BPS-D4}) with $L_1=L_2=0$ from a solution of the BPS system in \cite{Maldacena:2000mw}. Indeed, let $\tau(z)$ be 
the function that solves the following differential equation:
\begin{equation} 
\frac{d^2 \tau}{dz^2} + 2 \left(\frac{d\tau}{dz}\right)^2 = 2\,  \frac{d\tau}{dz} \,  
\left[4 e^{2\tau} -\frac{1}{4 z}\right]  \, .
 \label{master-equation-D4}
\end{equation}
It turns out that we can write a solution of the unflavored BPS equations in terms of this function $\tau(z)$. First of all, one can verify that the PDE \eqref{flavor-PDE-D4} with $L_1=L_2=0$ is solved by  a  $z(\rho,\sigma)$ that is given implicitly by means of the following equation:
\begin{equation}
\frac{\rho^2}{e^{4\tau}} + \frac{\sigma^2}{\frac{1}{2}\,e^{2\tau}}\,\frac{1}{\sqrt{z}\frac{d\tau}{dz}}=
\frac{1}{4m^2}\, . 
\label{implicit-D4}
\end{equation}
Moreover, the warp factor $H$ is given  in terms of $z(\rho,\sigma)$ and $\tau(z)$ by:
\begin{equation}
H= \frac{\sqrt{z}}{m^2 \left[4\sigma^2e^{2\tau}+ \rho^2 \sqrt{z}\left(\frac{d\tau}{dz}\right)^2 
e^{-2\tau}\right]}\, \, , 
\label{unflavor-H-D4}
\end{equation}
while the functions $g_1$ and $g_2$ are:
\begin{equation}
g_1= -\frac{\rho\, H}{64 m^2}\, \left(\frac{d\tau}{dz}\right)^2\,\left[1+ \frac{16 m^2 \sigma^2}
{\sqrt{z}\big(\frac{d\tau}{dz}\big)^2}\,-\,{4m^2\sigma^2\, e^{-2\tau}\over \sqrt{z}\,
{d\tau\over d z}}
\right]\,\,, \quad \quad \quad \quad g_2= -\frac{\rho\, e^{-2\tau}}{16m^2} \, \, . 
\label{unflavor-g1-g2-D4}
\end{equation}
One can easily show that the RR flux for the solution (\ref{unflavor-g1-g2-D4}) is correctly quantized when the constant $m$ is given by (\ref{m-D4}).

\begin{figure}
\centering
\includegraphics[scale=0.7]{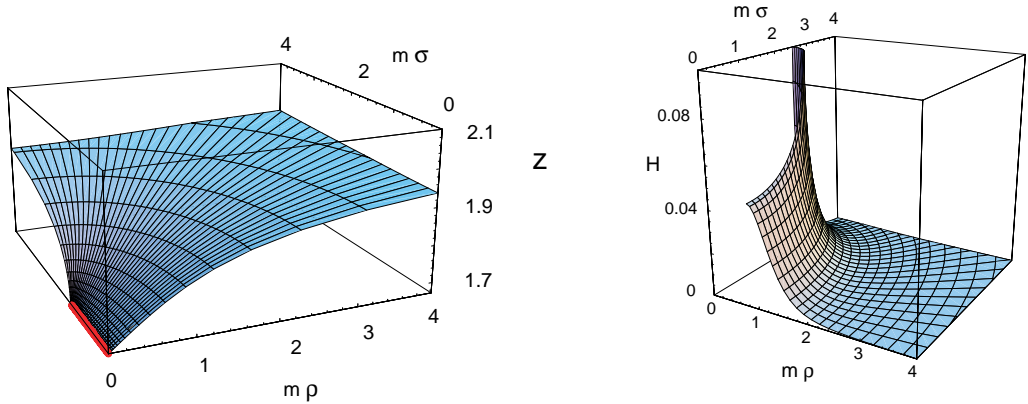}
\caption{Plots of $z$ and $H$ for the unflavored background of D4-branes. We are taking the UV parameters $z_*=2$ and $c=1$. On the right plot we only show the region  in which the warp factor $H$ is monotonic.}
\label{zHunflavoredD4}
\end{figure}

It is now straightforward to solve numerically first (\ref{master-equation-D4}) for $\tau(z)$ and use this result in the implicit  equation (\ref{implicit-D4}) to get $z(\rho, \sigma)$. The results are plotted in figure \ref{zHunflavoredD4} for $z(\rho,\sigma)$ and $H(\rho,\sigma)$. They are very similar to the ones found in section \ref{D3-section} for the wrapped D3-branes and all the comments made in subsection \ref{numerical-unflavored-D3} also apply to the present case. In particular, the function $H$ is not monotonic in the IR and one should only trust the region in which $H$ decreases as we move towards the UV. 

\subsubsection{The approximate unflavored solution in the UV}
\label{UV-D4}

When $\tau \rightarrow \infty$, $z$ approaches a constant value and, therefore,  the term $1/z$ on the right-hand side  of \eqref{master-equation-D4} becomes negligible compared to $e^{2\tau}$. Thus,approximately,  the master equation \eqref{master-equation-D4} becomes:
\begin{equation} 
\frac{d^2 \tau}{dz^2} + 2 \left(\frac{d\tau}{dz}\right)^2  \simeq  8 e^{2\tau}  \frac{d\tau}{dz}   \,\,,  \quad \quad  \quad \quad  (\tau \rightarrow \infty) \, .
\label{approx-UV-D4}
\end{equation}
This equation can be integrated once to give: 
\begin{equation}
\frac{d\tau}{dz} = 2 \Big(\,e^{2\tau} + c\, e^{-2\tau}\,\Big) \, , 
\end{equation}
while an additional integration gives $\tau$ as a function of $z$, namely:
\begin{equation}
e^{2\tau}\, = \, \sqrt{c}\cot \left[4 \sqrt{c} \left(z_{*} - z\right)\right] \, , 
\label{tau-z-UV-D4}
\end{equation}
with $c$ and $z_{*}$  being constants.  Notice also that, for large $\tau$, we can take
at leading order 
$\frac{d\tau}{dz} \simeq  2 e^{2\tau} $ and the implicit equation  \eqref{implicit-D4} becomes:
\begin{equation}
\rho^2\,+\,\frac{\sigma^2}{\sqrt{z}}\,=\,\frac{1}{4 m^2}\,\frac{1}{16 (z-z_{*})^2} \, .
\label{implicit-UV-D4}
\end{equation}
Eqs. (\ref{tau-z-UV-D4}) and (\ref{implicit-UV-D4}) show that for large $\tau$ one approximately has $z\approx z_*$ and we are far from the origin in the $(\rho,\sigma)$ plane. Moreover, by performing  an expansion around $z_{*}$  in (\ref{implicit-UV-D4}) and  by keeping only the first non-trivial term we have: 
\begin{equation} \label{UV-z}
z(\rho, \sigma) \thickapprox z_{*} - \frac{1}{8 m} \frac{z_{*}  ^{1/4}}{\sqrt{ 
\rho^2 z_{*}^{1/2} +\sigma^2}} \ .
\end{equation}
Notice also that, in the far UV where  $z\approx z_*$ , one has:
\beq
\sigma^2\,+\,\sqrt{z_*}\,\,\rho^2\approx {\sqrt{z_*}\over 4m^2}\,\,e^{4\tau}\,\,.
\eeq
Using these approximate UV expressions in 
 \eqref{unflavor-H-D4}  we arrive at the following UV estimate of the warp factor $H$:
\begin{equation} \label{UV-H}
H \approx \left(\frac{1}{2m}\right)^3 
\frac{ z_{*} ^{3/4}}{\left[\sqrt{\rho^2 z_{*}^{1/2} +\sigma^2}\, \right]^3}\, ,
\end{equation}
whereas from (\ref{unflavor-g1-g2-D4}) we get that $g_1$ and $g_2$ are approximately  given by:
\begin{equation} 
g_1\approx -{1\over 32 m^2}\,\,{\rho\,z_*^{{1\over 4}}\over 
\sqrt{\rho^2\,\,z_*^{1/2}+\,\sigma^2}}\,\Big[\,1\,+\,{1\over 2}\,
{\sigma^2\over \sqrt{z_*}\,\,\rho^2\,+\,\sigma^2}\,\Big]
\,\,,
\quad  \quad \qquad
g_2\approx -\frac{\rho z_{*}^{1/4}}{32 m^3 \sqrt{\rho^2\, z_{*}^{1/2} +\sigma^2}} \, \, . 
\label{UV-g1-g2}
\end{equation}
The above analysis suggests that in the UV region the combination $\rho^2 z_{*}^{1/2} +\sigma^2$, appearing in the denominator of both $z$ and $H$, plays a significant role. Having this in mind let us  define a new set of variables $u$ and ${\alpha}$ as 
in (\ref{u-alpha}) but now with $0\le\alpha\le \pi$.  In general, $z$ and $H$ depend on both coordinates, $u$ and ${\alpha}$, but in the UV  region  the ${\alpha}$ dependence disappears. Actually,  their expressions when $u\to\infty$ are:
\begin{equation}
z \rightarrow  z_{*}\,\,,\quad \ \quad \qquad
H^{1/2} \rightarrow
\frac{1}{2\sqrt{2} } \left(\frac{z_{*}^{1/4}}{m\, u}\right)^{3/2} \ .
\label{UV-functions}
\end{equation}
Using these values in the metric ansatz (\ref{D4-metric})  we end up with the following expression:
\begin{eqnarray} \label{UV}
ds^2_{UV} & \approx & 2\sqrt{2}
\left(\frac{m u}{z_{*}^{1/4} }\right)^{3/2} \, \left[dx^2_{1,2}+\frac{z_{*}}{m^2} \,d\Omega_2^2\right] + 
\frac{1}{2\sqrt{2} \,z_{*}^{1/8}}
\frac{du^2}{\left(m u\right)^{3/2}} + \nonumber \\
&+& \frac{1}{2\sqrt{2}\, z_{*}^{1/8}}\, {u^{1/2} \over m^{3/2}} \,
\left[d{\alpha}^2 +\frac{1}{4}\, \sin^2{\alpha}\,
\left[ (w^1)^2 + (w^2)^2 + \left(w^3 + \cos \theta d\phi \right)^2\right]
\right] \ ,
\end{eqnarray}
where $d\Omega_2^2\equiv d\theta^2+\sin^2 \theta\,d\phi^2$ is the line element of the $(\theta, \phi)$ two-sphere. In order to interpret the meaning of the results just found, let us recall that, given a background of type IIA theory such as the one we are considering, one can generate a solution of eleven-dimensional supergravity by uplifting the metric in terms of the standard  formula:
\begin{equation}
ds_{11}^2\,=\, e^{-{2\over 3}\Phi}\,ds^2_{10}\,+\,e^{{4\over 3}\Phi}\,(dx^3)^2\,\,,
\label{ds-11d}
\end{equation}
where $x^3$ is the eleventh M-theory coordinate. We shall apply (\ref{ds-11d}) to the ten-dimensional UV metric and dilaton written in eqs. (\ref{UV}) and (\ref{UV-functions}). After changing the radial variable $u$ by a new coordinate $y$, defined as:
\begin{equation}
y^2\,=\,\frac{2 m}{z_{*}^{1/4}}\, u\,,
\end{equation}
the resulting eleven-dimensional UV metric becomes:
\begin{eqnarray}
ds_{11}^2\,& \approx &\, y^2 \,
 \,\Big[\,dx^2_{1,3}\,+\,
{z_{*} \over m^2}\,d \Omega_2^2\,\Big]\,+\, \frac{1}{m^2}\, \Big({dy\over y}\Big)^2\,+\,\rc\rc
&+& {1 \over 4 m^2}\,
\left[d{\alpha}^2 +\frac{1}{4}\, \sin^2{\alpha}\,
\left[ (w^1)^2 + (w^2)^2 + \left(w^3 + \cos \theta d\phi \right)^2\right]\right]\,,
\label{AdS7-S4}
\end{eqnarray}
where $dx^2_{1,3}\,=\,dx^2_{1,2}+(dx^3)^2$. From (\ref{AdS7-S4}) we conclude that the uplifted metric is of 
the form $AdS_7\times S^4$, with the $AdS_7$ having two of its directions compactified in a two-sphere  and with the $S^4$ being  fibered over this $S^2$. Notice that the radius of the $AdS_7$ ($S^4$ ) factor in the asymptotic metric (\ref{AdS7-S4}) is $1/m$ ($1/(2m)$).  These results are, of course, consistent with the origin of the solution of \cite{Maldacena:2000mw}, as coming from M5-branes wrapping a two-cycle.

\subsection{Probe analysis}
In complete analogy with the analysis performed in section \ref{probe-D3}, we can test our background by using a color D4-brane extended along the Minkowski directions and those of the two-cycle. Again, the equilibrium position of such a brane occurs at the no-force SUSY locus $\sigma=0$. The coupling of the Yang-Mills theory living on the color branes can be obtained by introducing a worldvolume gauge field along the unwrapped directions and  expanding at second order in the gauge field strength. One gets:
\beq
{1\over g_{YM}^2}\,=\,{z(\rho,\sigma=0)\over 2\pi g_s\,m^2\sqrt{\alpha'}}\,\,.
\label{YM-D4}
\eeq
From our numerical results of figure \ref{zHunflavoredD4} we notice that $z(\rho,\sigma)$ grows monotonically as $\rho$ increases and reaches a constant asymptotic value when $\rho\to\infty$. Using this result we conclude that $g_{YM}$ decreases as we go to the UV. We can have an analytic estimate of this behavior by using the asymptotic results of subsection \ref{UV-D4}. Indeed, plugging (\ref{UV-z})  in (\ref{YM-D4}) we arrive at:
\beq
{1\over g_{YM}^2}\,\approx\,
{1\over g_{*}^2}\,-\,{1\over 16\pi g_s\,m^3\sqrt{\alpha'}}\,\,{1\over \rho}\,\,,
\label{YM-D4-UV}
\eeq
where  $g_{*}$ is the asymptotic value of the YM coupling ($g_{*}^2=2\pi g_s\,m^2\sqrt{\alpha'}/z_{*}$). Again, the behavior  displayed by our holographic model coincides qualitatively with the one found in the perturbation theory of the gauge theory dual. Indeed, if one relates the coordinate $\rho$ with the energy scale of the field theory by means of the naive relation $\rho=2\pi\alpha'\mu$, one gets that the right-hand side of (\ref{YM-D4-UV}) runs as $\mu^{-1}$, which is the same running obtained with the one-loop beta function (although the numerical coefficients are different). 

\subsection{Flavored solutions}

Let us now analyze the system (\ref{flavor-BPS-D4}) in the flavored case. According to our macroscopic analysis, the source  five-form $\Omega$ is parametrized by two functions $L_1$ and $L_2$, which must satisfy the consistency equation (\ref{L12-eq-D3}).  However, as was the case for the D3-brane system of section \ref{D3-section}, there is a simple solution to (\ref{L12-eq-D3}), namely (\ref{L12-D3-simple}), which leads to physically sensible results. Therefore, from now on we will assume that $L_1$ and $L_2$ are given  by the expressions displayed in (\ref{L12-D3-simple}). The corresponding smearing form $\Omega$ is:
\beq
\Omega\,=\,{C\over 2 \kappa_{10}^2\,T_4}\,\delta(\rho-\rho_Q)\,
d\rho\wedge  \Omega_2\,\wedge \Big[\,\sigma^2\,w^1\wedge w^2\,+\,
2\sigma\,d\sigma\wedge w^3\,\Big]\,\,,
\label{Omega-special-D4}
\eeq
where $C$ is the constant appearing in the expression of $L_2$ in (\ref{L12-D3-simple}).
In terms of the $y^i$ coordinates defined in (\ref{y-def}), the form $\Omega$ can be written as:
\beq
\Omega\,=\,-{2C\over \kappa_{10}^2\,T_4}\,\,\delta(\rho-\rho_Q)\,
d\rho\wedge  \Omega_2\wedge
\big[\,dy^1\wedge dy^2\,+\,dy^3\wedge dy^4\,\big]\,\,.
\label{Omega-ys-D4}
\eeq

Let us now see how we can recover the expression (\ref{Omega-ys-D4}) of $\Omega$ from a microscopic calculation. 
According to the brane array written at the beginning of this section, the flavor D4-branes span a dimension-two surface in the normal bundle $N_4$. Actually, it is straightforward to verify that the family of embeddings (\ref{planes-D3}) are calibrated by the form ${\cal K}$ written in (\ref{K-D4}) and, therefore, they preserve the supersymmetries of the background. The corresponding charge distribution obtained by an homogeneous distribution of D4-branes embedded along the family (\ref{planes-D3}) and smeared over the $(\theta, \phi)$ two-sphere at a given  fixed value $\rho_Q$  of the transverse coordinate $\rho$ is given by:
\beq
\Omega\,=\,\delta(\rho-\rho_Q)\,d\rho\,\wedge\,
{\sin\theta\, d\theta\wedge d\phi\over 4\pi}\,\wedge \Gamma\,\,,
\label{Omega-microD4}
\eeq
where $\Gamma$ is the two-form of eqs. (\ref{Gamma-D3}) and (\ref{Gamma-ys}) that results after averaging over the planes (\ref{planes-D3}). Plugging (\ref{Gamma-ys}) into (\ref{Omega-microD4}) we immediately get (\ref{Omega-ys-D4}) if the constant $C$ is related to the density $n_f$ as:
\beq
C\,=\,-2\kappa_{10}^2\,T_4\,\,{n_f\over 32\pi}\,\,.
\eeq
By using this result, we can rewrite the PDE  (\ref{flavor-PDE-D4}) as:
\begin{equation} 
2\, z \, \sqrt{z} \, \left( \sigma \ddot{z} \, + \, \dot{z} \right)\, =\, \sigma \, \left( {z'}^2 -\, 2\, z\, z'' \right) \,-\,{\pi\, n_f\over 2 m N_c}\,\,
 \sigma\,z^{3/2}\, \, \delta(\rho-\rho_Q)  \, ,
\label{PDE-D4flavored-with-constant}
\end{equation}
while the discontinuity of $z'$ at $\rho=\rho_Q$ is given by:
\beq
z'(\rho_Q+\epsilon, \sigma)\,-\,z'(\rho_Q-\epsilon, \sigma)\,=\,
-{\p\, n_f\over 4 m\, N_c}\,\,\sqrt{z(\rho_Q, \sigma)}\,\,.
\label{jump-D4}
\eeq

\begin{figure}
\centering
\includegraphics[scale=0.7]{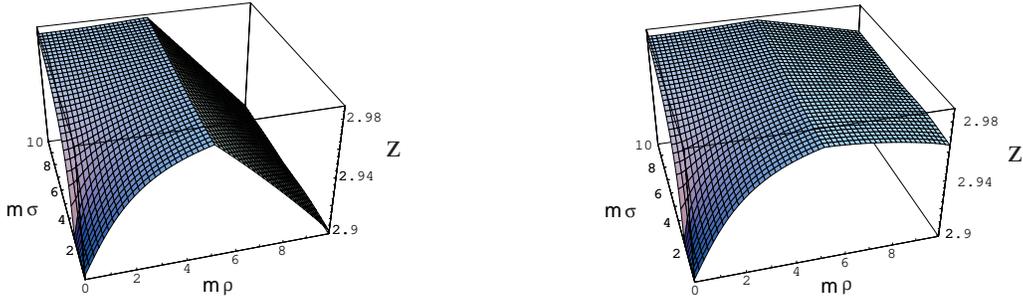}
\caption{Plots of $z(\rho,\sigma)$ obtained from the numerical integration of the PDE equation (\ref{PDE-D4flavored-with-constant}) for two different values of $x\equiv{\pi n_f\over m^2 N_c}$.  The left (right) plot corresponds to $x=1/5$ ($x=1/20$). In both cases we are taking $m\,\rho_Q=5$ and the $\rho<\rho_Q$ region has been obtained with the unflavored solution with UV parameters $c=-1$ and $z_*=3$.}
\label{zfl-D4}
\end{figure}

The PDE (\ref{PDE-D4flavored-with-constant}) with the jump condition (\ref{jump-D4}) can be integrated numerically by using the same strategy applied to the case of the D3-branes in section \ref{D3-section}. In figure \ref{zfl-D4} we have plotted the result of this numerical integration for two values of $n_f$. From these plots it is clear that the effect of the flavors in this solution is qualitatively the same as in the case of wrapped D3-branes and, therefore, the remarks made after (\ref{jump-D3}) are also applicable to the present case.

\section{ ${\cal N}=2$ 3d SQCD   from wrapped D5-branes}
\label{D5-section}

In this section we will study an alternative way of constructing a gravity dual of three-dimensional ${\cal N}=2$  gauge theories, namely by considering the background generated by D5-branes wrapping a three-cycle of a Calabi-Yau manifold of complex dimension three. In the unflavored case this setup was considered in refs. \cite{Gomis:2001aa,Gauntlett:2001ur}, where an explicit solution was found from gauged supergravity. 
The addition of flavor to this system was addressed in ref. \cite{Gaillard:2008wt} by means of a macroscopic approach similar to the one employed here. In the present paper we revisit this configuration with our methods and, in particular, we find a flavored solution with a good microscopic description in terms of a family of embeddings of D5-branes.

The brane setup we will analyze is represented by the following array:
\begin{center}
\begin{tabular}{|c|c|c|c|c|c|c|c|c|c|c|}
\multicolumn{3}{c}{ }&
\multicolumn{8}{c}
{$\overbrace{\phantom{\qquad\qquad\qquad\qquad\qquad}}^{\text{CY}_3}$}\\
\hline
&\multicolumn{3}{|c|}{$\mathbb{R}^{1,2}$}
&\multicolumn{3}{|c|}{$S^3$}
&\multicolumn{3}{|c|}{$N_3$}
&\multicolumn{1}{|c|}{$\mathbb{R}$}\\
\hline
$N_c$ D$5$ &$-$&$-$&$-$&$\bigcirc$&$\bigcirc$&$\bigcirc$
&$\cdot$&$\cdot$&$\cdot$&$\cdot$\\
\hline
$N_f$ D$5$ &$-$&$-$&$-$&$\bigcirc$&$\cdot$&$
\cdot$
&$-$&$-$&$\cdot$&$\cdot$\\
\hline
\end{tabular}
\end{center}
In order to formulate a specific ansatz for the metric and RR three-form, let us introduce some definitions.  Let $w^i$ be $SU(2)$ left-invariant one-forms. We shall consider a rotated version of these forms  by two angles $\theta$ and $\phi$, with $0\le\theta\le \pi$ and $0\le \phi <2\pi$.  Accordingly, we define three new one-forms  $S^i$ $(i=1,2,3)$ as:
\bear
&&
S^1=\cos\phi\,{w^1\over 2}-\sin\phi\,{w^2\over 2}\,,\rc\rc
&&
S^2=\sin\theta\,{w^3\over 2}-\cos\theta\left(\sin\phi\,{w^1\over 2}+
\cos\phi\,{w^2\over 2}\right)\,,\rc\rc
&&
S^3=-\cos\theta\,{w^3\over 2}-\sin\theta\left(\sin\phi\,{w^1\over 2}+
\cos\phi\,{w^2\over 2}\right)\,.
\label{rotomega}
\eear 
Similarly, we define two new one-forms $E^1$ and $E^2$, obtained from $d\theta$ and 
$\sin\theta d\phi$, that are the natural frame forms of a two-sphere,  as:
\bear
&&E^1=d\theta+\cos\phi\,{w^1\over 2}-\sin\phi\,{w^2\over 2}\,
=\,d\theta+S^1\,\,,
\rc\rc
&&E^2=\sin\theta\left(d\phi+\,{w^3\over 2}\right)-\,
\cos\theta\left(\sin\phi\,{w^1\over 2}+\cos\phi\,{w^2\over 2}\right)\,=\,
\sin\theta d\phi+\,S^2\,\,.
\label{Es}
\eear
The prototypical example of a  non-compact Calabi-Yau threefold with a three-cycle blown up is the deformed conifold. It turns out \cite{Gimon:2002nr} that the metric of the deformed conifold can be written in terms of the one-forms $S^i$ and $E^j$ as follows:
\bear
&&ds^2_6\,=\,{1\over 2}\,\,\mu^{{4\over 3}}\, K(\sigma)\,\Bigg[\,{1\over 3  K(\sigma)^3}\,
\Big((d\sigma)^2+4(S^3)^2\Big)\,+\,2\cosh^2\Big({\sigma\over 2}\Big)\,
\Big((S^1)^2+(S^2)^2\Big)\,+\,\rc\rc
&&\qquad\qquad\qquad\qquad+\,
\,2\sinh^2\Big({\sigma\over 2}\Big)\,\Big((E^1)^2+(E^2)^2\Big)\,\Bigg]\,\,,
\label{metric-conifold}
\eear
where $0\le\sigma<+\infty$ is the radial coordinate, $\mu$ is the deformation parameter  of the conifold and $K(\sigma)$ is the function:
\beq
K(\sigma)\,=\,{\big(\sinh (2\sigma)\,-\,2\sigma\,\big)^{{1\over 3}}\over 2^{{1\over 3}}\,\sinh \sigma}\,\,.
\eeq
We will take the metric (\ref{metric-conifold}) as the starting point to formulate our ansatz for the ten-dimensional metric associated to our brane setup. First of all, we add three Minkowski coordinates $x^{0,1,2}$ and a new non-compact coordinate $\rho$ parametrizing the transverse $\mathbb{R}$ ($-\infty<\rho<+\infty$). As before, we will parametrize the size of the three-cycle by a function $z=z(\rho,\sigma)$ and we will introduce a squashing between the three-cycle and the $S^2$ fiber. Accordingly, the string  frame metric takes the form:
\beq
ds^2=e^{\Phi}\left(dx_{1,2}^2+{z\over m^2}\,d\Omega_3^2\right)+
e^{-\Phi}\left\{{1\over z}\Big[d\sigma^2+\sigma^2\left[(E^1)^2+(E^2)^2\right]\Big]
+d\rho^2\right\}\,,
\label{D5-metric}
\eeq
where $\Phi=\Phi(\rho,\sigma)$ is the dilaton and $m^2$ is  now:
\beq
m^2={1\over g_s\,N_c\,\alpha'}\,. 
\label{mvalue}
\eeq
In eq. (\ref{D5-metric}) $d\Omega_3^2$ is the metric of the three-sphere, namely:
\beq
d\Omega_3^2\,=\,\sum_i\,(S^i)^2\,=\,{1\over 4}\,\sum_i\,(w^i)^2\,\,.
\eeq
The background is also endowed with an  RR three-form $F_3$ which, in complete analogy with the results of sections \ref{D3-section} and \ref{D4-section},  we will represent in terms of  two-forms $C_2$ and $\Lambda$ as:
\beq
F_3\,=\,d\,C_2\,+\,\Theta(\rho-\rho_Q)\,d\Lambda\,\,.
\label{F3-D5}
\eeq
We will adopt the following ansatz for $C_2$ and $\Lambda$:
\bear
&&C_2\,=\,g_1\,E^1\wedge E^2\,+\,g_2\,S^1\wedge S^2\,\,,\qquad\qquad
g_i\,=\,g_i(\rho, \sigma)\,\,,\rc\rc
&&\Lambda\,=\,L_1\,E^1\wedge E^2\,+\,L_2\,S^1\wedge S^2\,\,,\qquad\qquad
L_i\,=\,L_i(\sigma)\,\,. 
\label{C2-Lambda-D5}
\eear
Notice that the two-form $\Lambda$ introduces a violation of the Bianchi identity of the form:
\beq
dF_3\,=\,\delta(\rho-\rho_Q)\,d\Lambda\,\,,
\label{Bianchi}
\eeq
which corresponds to sources localized at a fixed value $\rho_Q$ of the $\rho$ coordinate. Let us now write down the explicit expression for $F_3$.
Since we have:
\beq
d\big(E^1\wedge E^2\big)\,=\,-d\big(
{ S}^1\wedge { S}^{2}\big)\,=\,E^1\wedge \,{\ S}^1\wedge { S}^{3}\,+\,
E^2\wedge\,{ S}^2\wedge { S}^{3}\,\,.
\label{dE1E2}
\eeq
Then,  the RR three-form $F_3$ given by (\ref{F3-D5}) and (\ref{C2-Lambda-D5}) takes the form:
\bear
&&F_3\,=\,\Big(\,dg_1\,+\,\Theta(\rho-\rho_Q)\,\dot L_1\,d\sigma\,\Big)\,\wedge
E^1\wedge E^2\,+\,
\Big(\,dg_2\,+\,\Theta(\rho-\rho_Q)\,\dot L_2\,d\sigma\,\Big)\,\wedge
S^1\wedge S^2\,+\,\rc\rc
&&\qquad
+\,\Big(g_1-g_2\,+\,\Theta(\rho-\rho_Q)(L_1-L_2)\,\Big)\,
\Big(\,E^1\wedge S^1\wedge S^3\,+\,E^2\wedge S^2\wedge S^3\,\Big)\,\,.
\label{F3-explicit-D5}
\eear
It is natural to consider the following basis of vielbein one-forms:
\bear
&&e^i=e^{\Phi\over2}\,dx^i\,,\quad(i=0,1,2)\,,\qquad
e^j=e^{\Phi\over2}\,{\sqrt{z}\over m}\,S^{j-2}\,,\quad(j=3,4,5)\,,\qquad
e^6={e^{-{\Phi\over2}}\over\sqrt{z}}\,d\sigma\,,\rc\rc
&&
e^7=e^{-{\Phi\over2}}\,{\sigma\over \sqrt{z}}\,E^1\,,\qquad
e^8=e^{-{\Phi\over2}}\,{\sigma\over \sqrt{z}}\,E^2\,,\qquad
e^9=e^{-{\Phi\over2}}\,d\rho\,.
\label{D5-frame}
\eear

In order to perform the supersymmetry analysis of the ansatz (\ref{D5-metric})-(\ref{F3-explicit-D5}) in type IIB supergravity, let us impose the following set of projections to the Killing spinors:
\beq
\Gamma_{3478}\,\epsilon=\epsilon\,,\qquad
\Gamma_{3568}\,\epsilon=\epsilon\,,\qquad
\Gamma_{6789}\,\epsilon=\,-\tau_1\,\epsilon\,,
\label{D5-projs}
\eeq
which leave four supercharges unbroken. In (\ref{D5-projs}) $\tau_1$ is the first Pauli matrix. The corresponding BPS equations are:
\bear
&&\dot{\Phi}=-m^2\,e^{-2\Phi}\,{\sigma\over2z^2}-\,e^{2\Phi}\,{\sqrt{z}\over2\sigma^2}\,g_1'\,,\rc \rc
&&\Phi'={1\over2\sqrt{z}}\left[e^{2\Phi}\,{z^2\over\sigma^2}\, 
\big(\,\dot{g_1}+\Theta(\rho-\rho_Q)\,\dot L_1\,\big)\,
+\,{3m^2\over\sigma}\big(g_1-g_2+\Theta(\rho-\rho_Q)(L_1-L_2)\big)\right]\,,\rc \rc
&&\dot z=2m^2\,e^{-2\Phi}\,{\sigma\over z}\,,\rc\rc
&& z'=2m^2\,{\sqrt{z}\over\sigma}\big(g_2-g_1
+\Theta(\rho-\rho_Q)(L_2-L_1)\big)\,,\rc\rc
&&\dot{ g_2}=-{1\over\sigma}(g_1-g_2)\,-\,
\Theta(\rho-\rho_Q)\,\Big[\,\dot L_2\,-\,{L_2-L_1\over \sigma}\,\Big]\,\,,\rc\rc
&&g_2'=-\,e^{-2\Phi}\,{\sigma\over\sqrt{z}}\,.
\label{BPS-D5}
\eear
The consistency of this system requires that the $L_i$'s satisfy the condition:
\beq
\dot L_2\,=\,{L_2-L_1\over \sigma}\,\,.
\label{L1L2eq-D5}
\eeq
Moreover,  from the BPS system (\ref{BPS-D5}) we get the following PDE for the function $z(\rho,\sigma)$:
\beq
z\left(z\,\ddot z+{1\over2}\,\dot z^2\right)={1\over 2}\,z'^{\,2}-\,z\,z''\,+\,
{2 m^2\,(L_2-L_1)\over\sigma}\,z^{3/2}\,\delta(\rho-\rho_Q)\,.
\label{pdeflavD5}
\eeq
One can check that, if eqs. (\ref{BPS-D5}) and (\ref{L1L2eq-D5}) hold, then the equations of motion of the gravity plus smeared branes system are fulfilled.  Notice the difference between the consistency conditions (\ref{L1L2eq-D5}) and (\ref{L12-eq-D3}), which is due to the fact that in this D5-brane case the flavor branes are codimension one in the normal bundle $N_3$, whereas in the setups of sections \ref{D3-section} and \ref{D4-section} the flavor branes span a codimension two submanifold of $N_4$.

The generalized  calibration form for this case is a six-form of the type:
\beq
{\cal K}\,=\,{1\over 6!}\,{\cal K}_{a_1\cdots a_6}\,e^{a_1\,\cdots a_6}\,\,.
\label{K-def-D5}
\eeq
The different components of ${\cal K}$  are given by:
\beq
{\cal K}_{a_1\cdots a_6}\,\equiv\,-e^{-{\Phi\over 2}}\,\,\epsilon^{\dagger}\,\Gamma_{a_1\,\cdots a_6}\,\tau_1\,\epsilon\,\,,
\label{K-bilinear}
\eeq
where $\epsilon$ is a Killing spinor of the background, normalized as 
$e^{-{\Phi\over 2}}\,\,\epsilon^{\dagger}\,\epsilon\,=\,1$, and the minus sign in the definition (\ref{K-bilinear}) has been introduced for convenience.  By using the SUSY projections satisfied by our solutions (eq. (\ref{D5-projs})), we get the actual components of ${\cal K}$  in the frame basis (\ref{D5-frame}), namely:
\beq
{\cal K}\,=\,-e^{012}\wedge\left(e^{345}+e^{367}+e^{468}-e^{578}\right)\,\,.
\label{K-D5}
\eeq
The BPS system (\ref{BPS-D5}) can be now summarized in the following equations satisfied by the calibration form ${\cal K}$:
\beq
{}^*\,F_3\,=\,-d\big(\,e^{-\Phi}\,{\cal K}\,\big)\,\,,\qquad\qquad
d\big({}^*\,{\cal K})\,=\,0\,\,.
\label{calibration-conditions-D5}
\eeq

\subsection{The unflavored solution}
The background of refs. \cite{Gomis:2001aa,Gauntlett:2001ur}, obtained in seven-dimensional gauged supergravity and afterwards uplifted to ten dimensions, provides a solution of the BPS system (\ref{BPS-D5}) and of the PDE equation (\ref{pdeflavD5}) in the unflavored case $L_1=L_2=0$. Let us present this solution, entirely written in our variables. The function $z(\rho,\sigma)$ is given as the solution of the implicit equation:
\beq
{\sigma^2\over z^{3/2}\left(I_{3/4}({z\over 2})-c\,K_{3/4}({z\over 2})\right)^2}+{\rho^2\over\sqrt{z}\left(I_{-1/4}({z\over 2})+c\,K_{1/4}({z\over 2})\right)^2}={1\over 4\,m^2}\,,
\label{impz}
\eeq
where $I_{\nu}$ and $K_{\nu}$ are modified Bessel functions. One can, indeed, check that the function  $z(\rho,\sigma)$ defined by (\ref{impz}) solves the PDE equation (\ref{pdeflavD5}) for $L_1=L_2=0$. Let us next define the function $x(z)$ as:
\beq
e^{x(z)}\,\equiv\,\Bigg[\,{
I_{-1/4}({z\over 2})+c\,K_{1/4}({z\over 2})\over
I_{3/4}({z\over 2})-c\,K_{3/4}({z\over 2})}\,\Bigg]^{1\over 2}\,\,.
\eeq
Then, the dilaton $\Phi(\rho,\sigma)$ is given by:
\beq
e^{\Phi}\,=\,m\,\sqrt{e^{-6x}\,\rho^2\,+\,z^{-1}\,e^{2x}\,\sigma^2}\,\,.
\eeq
In order to write down the functions $g_1$ and $g_2$ that parametrize the three form $F_3$, let us now define the angle $\psi$ ($0\le \psi\le \pi$) as:
\beq
\cot\psi\,=\,z^{{1\over 2}}\,e^{-2x}\,{\rho\over \sigma}\,\,.
\eeq
In terms of $\psi$ the functions $g_1$ and $g_2$ are  simply given by:
\beq
g_1\,=\,{\psi\over m^2}\,-\,e^{-2\Phi-4x}\,{\rho\,\sigma\over\sqrt{z}} \,\,,\qquad\qquad
g_2\,=\,{\psi\over m^2}\,\,.
\eeq
This unflavored solution was analyzed in detail in ref. \cite{Gauntlett:2001ur}, where it was argued that, for $c>0$, it has a good  IR singularity corresponding to a linear distribution of fivebranes. By means of a probe computation, it was proposed in \cite{Gauntlett:2001ur} that the solution describes a slice of the Coulomb branch of ${\cal N}=2$ , $d=3$ pure YM theory. Actually, one can also prove in this case that the SUSY locus of color D5-brane probe occurs at $\sigma=0$ and that the Yang-Mills coupling is related to $z(\rho,\sigma=0)$ by:
\beq
{1\over g_{YM}^2}\,=\,{\big[\,z(\rho,\sigma=0)\,\big]^{{3\over 2}}\over (4\pi)^2\,g_s\,\alpha'\,m^3}\,\,.
\label{gYM-D5}
\eeq
Notice that now $z$ does not approach a constant value in the UV. Indeed, from (\ref{impz}) one gets that it grows logarithmically as $z(\rho,0)\sim 2\log (m\rho)$ for large $\rho$. Accordingly, $g_{YM}$ in (\ref{gYM-D5}) decreases as a power of a logarithm of $\rho$ in the UV, contrary to the result expected  for a three-dimensional gauge theory. Notice also that the dilaton blows up in the UV and, therefore, a string theory completion is needed in this region.

\subsection{Flavored solutions}
The modification of the Bianchi identity of $F_3$ induced by a set of flavor D5-branes distributed with a charge density four-form $\Omega$ is:
\beq
dF_3\,=\,2\kappa_{10}^2\,T_5\,\Omega\,\,.
\label{dF3-Omega-D5}
\eeq
By comparing (\ref{dF3-Omega-D5}) with the value of $dF_3$ for our ansatz (eq. (\ref{Bianchi})), one immediately gets the smearing form $\Omega$ in terms of $\Lambda$, namely:
\beq
\Omega\,=\,{\delta(\rho-\rho_Q)\,d\rho\,\wedge d\Lambda\over 2\kappa_{10}^2\,T_5}\,\,.
\eeq
As in the D3- and D4-brane cases studied above, we will concentrate on studying the configuration in which the function $L_1$ that parametrizes $\Lambda$ vanishes. In this case the compatibility condition (\ref{L1L2eq-D5}) can be easily solved:
\beq
L_1\,=\,0\,,\qquad L_2\,=\,A\,\s\,,
\eeq
with $A$ being a constant. The smearing form for this configuration can be straightforwardly obtained by using (\ref{dE1E2}) to compute  $d\Lambda$. One gets:
\beq
\Omega=\frac{A\,\delta(\rho-\rho_Q)}{2\,\kappa_{10}^2\,T_5}\td\r\wedge\left(\td\s\wedge S^1\wedge S^2-\s\left(S^1\wedge S^3\wedge E^1+S^2\wedge S^3\wedge E^2\right)\right).
\label{eqn:Omega3dugly}
\eeq
If we now substitute in \eqref{eqn:Omega3dugly} the expression for the one-forms $S^{1,2,3}$ and $E^{1,2}$ written in eqs. (\ref{rotomega}) and (\ref{Es}), we see that they can be rearranged so that $\Omega$ can be recast as:
\begin{align}
\Omega=-\frac{A\,\delta(\rho-\rho_Q)}{8\,\kappa_{10}^2\,T_5}\td\r\wedge &\left[\td\left(\s\sin\th\sin\f\right)\wedge w^2\wedge w^3+\td\left(\s\sin\th\cos\f\right)\wedge w^3\wedge w^1+\right.\nn\\
&\;+\left.\td\left(\s\cos\th\right)\wedge w^1\wedge w^2\right].
\label{eqn:Omega3d}
\end{align}
Let us introduce now the following set of cartesian coordinates for $N_3$:
\beq
y^1\,=\,\sigma\sin\theta\sin\phi\,\,,\qquad
y^2\,=\,\sigma\sin\theta\cos\phi\,\,,\qquad
y^3\,=\,\sigma\cos\theta\,\,.
\label{y-coordinates-D5}
\eeq
In  terms of these cartesian coordinates the smearing form has the following appealing  and compact form:
\beq
\Omega=-\frac{A\,\delta(\rho-\rho_Q)}{16\,\kappa_{10}^2\,T_5}\,\,
\epsilon^{ijk}\,d\rho\wedge
dy^i\wedge w^j\wedge w^k\,\,.
\label{Omega-dys-D5}
\eeq
In the next subsection we will identify a set of supersymmetric embeddings which give rise to a charge density distribution given by (\ref{Omega-dys-D5}).

\subsection{Microscopic analysis}
First of all, let us determine a continuous set of supersymmetric D5-brane embeddings distributed as the flavor branes of our setup.  In these embeddings  the D5-brane is located at a fixed $\rho=\rho_Q$, is extended along a two-dimensional plane in $N_3$ and, simultaneously,  wraps an $S^{1}$ inside the three-cycle $S^3$.  The particular $S^1\subset S^3$ that the brane wraps depends on the plane that it spans in $N_3$. In order to describe this relation, we first define the three-vector $\vec n$ as:
\beq
\vec n\,=\,(\sin\alpha\sin\beta\,,\,\sin\alpha\,\cos\beta\,,\,\cos\alpha\,)\,\,,
\eeq
where $0\le\alpha\le \pi$ and $0\le\beta\le 2\pi$. Let us next consider the family of embeddings in $N_3$ defined by  the following function:
\beq
f\,(\,\vec y\,)\equiv\,\vec n\cdot\vec y\,-\,y_{*}\,=\,0\,\,,
\eeq
where $\vec y\,=\,(y^1,y^2,y^3)$ are the cartesian coordinates defined in (\ref{y-coordinates-D5}). This is the equation of a plane having $\vec n$ as its normal direction and with $y_{*}$ being its minimal distance to the $\vec y=0$ origin of $N_3$.  Let us next define the following two tangent vectors to the plane:
\bear
&&\vec t_1\,=\,(\cos\beta\,,\,-\sin\beta\,,\,0)\,\,,\rc\rc
&&\vec t_2\,=\,(\cos\alpha\sin\beta\,,\,\cos\alpha\cos\beta\,,\,-\sin\alpha)\,\,.
\eear
Notice that $(\vec t_1\,,\,\vec t_2\,,\vec n)$ is an orthonormal basis in $\mathbb{R}^3$. 
Let us  now arrange the three $SU(2)$ one-forms $w^i$ in a vector, namely:
$\vec w\,=\,(w^1,  w^2,  w^3)$. If we define the one-forms:
\beq
w_{t_i}\,\equiv\,\vec t_i\cdot \vec w\,\,,\qquad\qquad
w_{n}\,\equiv\,\vec n\cdot \vec w\,\,,
\eeq
then, the embeddings in the $S^3$ are defined by the two differential conditions:
\beq
w_{t_1}\,=\,0\,\,,
\qquad\qquad
w_{t_2}\,=\,0\,\,.
\label{wt1=wt2=0}
\eeq
Notice that the system (\ref{wt1=wt2=0}) is integrable due to the property:
\beq
dw_{t_1}\,=\,-w_{t_2}\,\wedge w_{n}\,\,,\qquad\qquad
dw_{t_2}\,=\,- w_{n}\,\wedge w_{t_1}\,\,,
\eeq
which shows that the differential equations derived from (\ref{wt1=wt2=0}) are on involution and, according to Frobenius' theorem, they are integrable. Moreover, in order to show that the corresponding embedding preserves the supersymmetry of the background, one has to verify the fulfillment of the generalized calibration condition:
\beq
\hat {\cal K}\,=\,{\rm Vol}({\cal M}_6)\,\,,
\label{cal-cond-D5}
\eeq
where $\hat{\cal K}$ is the pullback of the six-form (\ref{K-D5}). To verify (\ref{cal-cond-D5}) we will come back for a while  to our original $(\sigma, \theta, \phi)$ coordinates of $N_3$. Notice, first of all, that we can use the differential equations (\ref{wt1=wt2=0}) to express the pullback of two of the $w^i$'s in terms of the third one. For example, if $\alpha\not=\pi/2$ we can write:
\beq
\hat w^1\,=\,\sin\beta\,\tan\alpha\,\,\hat w^3\,\,,\qquad\qquad
\hat w^2\,=\,\cos\beta\,\tan\alpha\,\,\hat w^3\,\,.
\eeq
Using this result we can compute the pullback of the different one-forms of the basis (\ref{D5-frame}) and, thus, one can compute the pullback of the calibration form ${\cal K}$. The result can be written as:
\beq
{\hat K}\,=\,{\sigma^2\,e^{\phi}\over 2 m^3\,\sqrt{2z}}\,\,
{\sec\alpha \,\sin\theta\over 
\cos\alpha\cos\theta+\cos(\beta-\phi)\,\sin\alpha\,\sin\theta}\,\,
dx^0\wedge dx^1\wedge d x^2\wedge \hat w^3 \wedge d\theta\wedge d\phi\,\,.
\eeq
One can also compute the induced metric and check that, indeed, eq. (\ref{cal-cond-D5}) is satisfied and, therefore, the embeddings  described above are supersymmetric, as claimed. 

Let us now verify that the charge distribution generated by this family of D5-brane configurations is just given by a six-form $\Omega$  such as the one written in (\ref{Omega-dys-D5}). Following the same methodology that we applied for the D3- and D4-brane cases,  let us write the smearing form $\Omega$ as:
\beq
\Omega\,=\,\delta(\rho-\rho_Q)\,d\rho\,\wedge \Gamma\,,
\label{Omega-initialmicro-D5}
\eeq
where $\Gamma$ is the following three-form:
\beq
\Gamma\,=\,\int d\mu\,\delta(f)\,\big[\,df\wedge \Gamma_{S^3}\,\big]\,\,,
\label{Gamma-intmu-D5}
\eeq
with the integration measure $d\mu$ given by:
\beq
d\mu\,=\,{\sin\alpha \,d\alpha\, d\beta\over 4\pi}\,\,n_f\,dy_*\,\,,
\label{dmu-D5}
\eeq
and  with $\Gamma_{S^3}$ being the charge density distribution obtained after averaging over the possible embbedings on the $S^3$ for a given plane in $N_3$. 
In (\ref{dmu-D5}) $n_f$ is a density of flavor branes.  The integral over $y_*$ in (\ref{Gamma-intmu-D5}) can be immediately performed by using $\delta(f)\,=\,\delta(\vec n\cdot\vec y\,-\,y_{*})$. One gets:
\beq
\Gamma\,=\,{n_f\over 4\pi}\,\,
\int \sin\alpha\,d\alpha\,d\beta\,\,\big[\,df\wedge \Gamma_{S^3}\,\big]\,\,.
\label{Gamma-dalpha-dbeta}
\eeq

In order to determine $\Gamma_{S^3}$  let us  first consider the particular plane with  $\alpha=\beta=0$. In this case the differential equations for the embedding in the $S^3$ are just $w^1=w^2=0$. After looking at the parametrization (\ref{w123}) of 
 $w^1$ and $w^2$ in terms of the three angles $(\hat \theta, \hat \phi, \hat \psi)$,  one immediately realizes that the equation for this embedding 
 can be integrated as:
\beq
f_{\phi}\equiv \hat\phi\,-\, \hat\phi_*\,=\,0\,\,,
\qquad\qquad
f_{\theta}\equiv \hat\theta\,-\, \hat\theta_*\,=\,0\,\,,
\eeq
where $\hat\phi_*$ and $\hat\theta_*$ are constant angles. 
These embeddings depend on two continuous parameters $\hat\phi_*$ and
$\hat\theta_*$ which span a two-sphere. Therefore the corresponding two-form 
$\Gamma_{S^3}$ is given by:
\beq
\Gamma_{S^3}\,=\,\int {\sin \hat\theta_{*}\,d \hat\phi_{*}\,d \hat\theta_{*}\over 4\pi}\,\,
\delta(f_{\phi})\,\delta(f_{\theta})\,\,
\big[\,df_{\phi}\wedge df_{\theta}\,\big]\,\,,
\eeq
where we are integrating over $\hat\phi_{*}$ and $\hat\theta_{*}$ with the invariant measure of the two-sphere. These integrals are immediately performed with the help of the Dirac $\delta$-functions:
\beq
\Gamma_{S^3}\,=\,{\sin \hat\theta\over 4\pi}\,\,d \hat\phi\wedge d \hat \theta\,=\,
{1\over 4\pi}\,\,w^1\wedge w^2\,\,.
\eeq
From this result it is clear that the generalization to arbitrary values of $\alpha$ and $\beta$ is:
\beq
\Gamma_{S^3}\,=\,{1\over 4\pi}\,
w_{t_1}\,\wedge
w_{t_2}\,\,.
\label{GammaS3-D5}
\eeq
Taking into account that $\vec n\,=\,\vec t_1\times \vec t_2$ or, in components, 
$n^i\,=\,\epsilon^{ijk}\,t_1^j\,t_2^k$, we can write:
\beq
w_{t_1}\,\wedge
w_{t_2}\,=\,{1\over 2}\,\epsilon^{ijk}\,n^i\,w^j\,\wedge\,w^k\,\,.
\eeq
Thus, we can rewrite (\ref{GammaS3-D5}) as:
\beq
\Gamma_{S^3}\,=\,{1\over 8\pi}\,\epsilon^{ijk}\,n^i\,w^j\,\wedge\,w^k\,\,,
\eeq
and, plugging this result in (\ref{Gamma-dalpha-dbeta}), we find that  $\Gamma$ can be represented as:
\beq
\Gamma\,=\,{n_f\over 32 \pi^2}\,\,
\Bigg[\,\int_{0}^{\pi}\,\sin\alpha\,d\alpha\,\int_0^{2\pi}\,d\beta\,\,n^i\,\,n^j\,\,\Bigg]\,\,
\epsilon^{jkl}\,dy^i\wedge w^k\wedge w^l\,\,.
\eeq
Moreover, using the fact that:
\beq
\int_{0}^{\pi}\,\sin\alpha\,d\alpha\,\int_0^{2\pi}\,d\beta\,\,n^i\,\,n^j\,=\,
{4\pi\over 3}\,\,\delta^{ij}\,\,,
\eeq
we get:
\beq
\Gamma\,=\,{n_f\over 24 \pi}\,\,\epsilon^{ijk}\,dy^i\wedge w^j\wedge w^k\,\,.
\eeq
Therefore,  by combining  this result with   (\ref{Omega-initialmicro-D5}) we conclude that the smearing form $\Omega$ for this family of embeddings  can be written as:
\beq
\Omega\,=\,{n_f\over 24 \pi}\,\epsilon^{ijk}\,\delta(\rho-\rho_Q)\,d\rho\wedge
dy^i\wedge w^j\wedge w^k\,\,.
\label{Omega-micro-D5}
\eeq
By comparing (\ref{Omega-micro-D5}) and the macroscopic expression of $\Omega$ written in  (\ref{Omega-dys-D5}) we get that both expressions coincide if  the constant $A$ is identified with:
\beq
A\,=\,-2\kappa_{10}^2\,T_5\,\,{n_f\over 3\pi}\,\,.
\eeq
Taking into account that $2\kappa_{10}^2\,T_5\,=\,4\pi^2\,g_s\,\alpha'$, we can rewrite the previous relation as:
\beq
A\,=\,-{4\pi\over 3}\,\,g_s\,\alpha'\,n_f\,\,.
\label{constantA-D5}
\eeq
\subsection{Numerical results}

\begin{figure}
\centering
\includegraphics[scale=0.7]{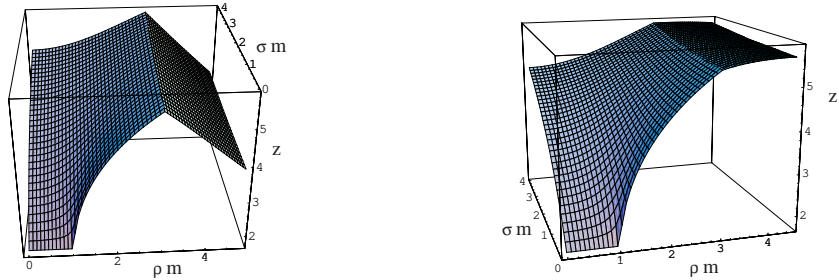}
\caption{Plots of $z(\rho,\sigma)$ obtained from the numerical integration of the PDE (\ref{pdeflavD5-with-constant}) for two values of $x\equiv {8\pi n_f\over 3 m N_c}$. The left curve corresponds to $x=2/3$, while the curve on the right was obtained for $x=1/5$. In these two cases $m\,\rho_Q=3$ and we are considering the case with $c=1$ for
$\rho<\rho_Q$, where $c$ is the constant appearing in the unflavored solution (\ref{impz}). }
\label{zflavoredD5}
\end{figure}

Using the expression (\ref{constantA-D5}) of the constant $A$, one can straightforwardly demonstrate that the PDE (\ref{pdeflavD5}) for the flavored system takes the form:
\beq
z\left(z\,\ddot z+{1\over2}\,\dot z^2\right)={1\over 2}\,z'^{\,2}-\,z\,z''\,-\,
{8\pi\,n_f\over 3 N_c}
\,z^{3/2}\,\delta(\rho-\rho_Q)\,.
\label{pdeflavD5-with-constant}
\eeq
The discontinuity of $z'$ at $\rho=\rho_Q$ in this case is given by:
\beq
z'(\rho_Q+\epsilon, \sigma)\,-\,z'(\rho_Q-\epsilon, \sigma)\,=\,-
{8\pi\,n_f\over 3 N_c}\,\sqrt{z(\rho_Q,\sigma)}\,\,.
\label{jump-D5}
\eeq
In figure \ref{zflavoredD5} we show the numerical result of the integration of the PDE (\ref{pdeflavD5-with-constant}) for $z(\rho,\sigma)$ which  reduces to the unflavored solution for $\rho<\rho_Q$ and such that its first partial derivative with respect to $\rho$ has the discontinuity given in (\ref{jump-D5}).

\section{Summary and conclusions}
\label{conclusions}

In this paper we have studied supergravity backgrounds which are obtained by wrapping D-branes on cycles of a Calabi-Yau cone. In their unflavored version these backgrounds were obtained some time ago by uplifting to ten dimensions the corresponding solutions in lower dimensional gauged supergravity. These uplifted solutions depend on several angles and are quite involved. Here we have rewritten them in terms of the two radial coordinates $(\rho,\sigma)$. In spite of the fact that one has to deal with functions depending on these two variables, this form of presenting these backgrounds is a simplification that clarifies their interpretation and allows to easily find their generalization when flavor sources are added. In the cases of the D3- and D4-branes, the solutions of the BPS systems in gauged supergravity of ref. \cite{Maldacena:2000mw} can be obtained by solving the master equations (\ref{master}) and (\ref{master-equation-D4}) and using these results in the implicit equations  (\ref{implicitD3}) and (\ref{implicit-D4}) respectively.  In the case of the D5-brane, the analytic solution of the unflavored background of refs. \cite{Gomis:2001aa,Gauntlett:2001ur} is obtained, in terms of our variables,  by solving the implicit equation (\ref{impz}).

The most important results of this paper are the construction of backgrounds which include the backreaction due to a large number of flavor branes. The latter act as extended dynamical sources for the different supergravity fields and, in particular, they modify the Bianchi identity of the RR fields. In order to deal with this problem,  we have first adopted an effective macroscopic approach in which the precise knowledge of the flavor brane embeddings is not needed. 

Using the preservation of supersymmetry and the compatibility with the metric ansatz as guiding principles, we were able to parametrize  the violation of the Bianchi identity induced by the flavor branes by means of two functions $L_1$ and $L_2$ and we argued, as in ref. \cite{Arean:2009gc}, that the most physically sensible configuration corresponds to taking $L_1=0$. We have verified this fact by reproducing the results found in the macroscopic approach by means of a careful analysis of the supersymmetric configurations of the flavor branes. We first found a family of embeddings which satisfy a generalized calibration condition and such that, in an appropriate system of coordinates, they can be regarded as planes in the directions orthogonal to the cycle wrapped by the color branes. We then computed the RR charge density $\Omega$ resulting after distributing homogeneously the flavor branes throughout the family of calibrated planes and we checked that the result coincides with the one found in the macroscopic approach. In turn, this result allowed us to relate the constants $A$ and $C$ of the macroscopic approach to the density $n_f$ of flavor branes. Once this identification is performed, the flavored backgrounds are obtained by integrating numerically the PDE equation for $z(\rho,\sigma)$ with the condition that the solution reduces to the unflavored one for $\rho<\rho_Q$ and such that the first derivative  $z'(\rho,\sigma)$ jumps at the location $\rho=\rho_Q$  of the flavor brane as dictated by the corresponding BPS equation.

Let us now comment on some points not addressed in our work that could be interesting to analyze in the  future. First of all, we could study the fluctuations of a flavor probe brane embedded along one of the supersymmetric planes. By studying the normalizable modes one would find the meson mass spectrum. The results should be close to the ones analyzed in refs. \cite{Angel,Arean:2008az,Ramallo:2008ew} for other backgrounds similar to the ones analyzed here. In particular, by comparing the spectra for the quenched and unquenched backgrounds one could study the effects of the backreaction on the meson masses. When the probe brane is located at $\rho=\rho_q\le \rho_Q$ the spectrum for the backreacted solution is the same as the one without backreaction since both metrics coincide  for $\rho\le\rho_Q$. On the contrary, for $\rho_q>\rho_Q$ this is not the case and, from the change in the mass spectra, one can extract, in the holographic setup,  the corrections  produced by charge screening effects due to quark loops. As in other similar cases analyzed before \cite{Angel,Arean:2008az,Ramallo:2008ew,BCPR}, one expects these screening effects to be small.

	In this paper we have only studied configurations of flavor branes extended along calibrated planes. There exist, however, more general embeddings that preserve the same supersymmetry (see, for example, section \ref{micro-D3} ). Generically, these other embeddings represent configurations in which the flavor branes end on the color branes in such a way that both types of branes are recombined. On the field theory side these solutions should correspond to Higgs branches of the corresponding supersymmetric theory.

One aspect that is missing for the kind of gravity duals analyzed here is the precise dictionary between them and the corresponding field theory. If we had this dictionary we could extract more precise information from our solutions. We could, for example, understand the relation between the field theory duals to the backgrounds of sections
\ref{D4-section} and \ref{D5-section}, which should correspond to two three-dimensional gauge theories with different UV completions. Nevertheless, we regard our work as part of the efforts devoted to approach the gauge/gravity duality to particle physics phenomenology. In the absence of a true gravity dual of QCD, the best that one can do is trying to uncover universal features of the duality by systematically analyzing different backgrounds and their generalizations. We hope that some of the results presented here could be useful in this endeavor.

\section*{Acknowledgments}
We are grateful to J\'er\^ome Gaillard, Carlos N\'u\~nez, \'Angel Paredes, Johannes Schmude, Jonathan Shock and Javier Tarr\'\i o for discussions and a critical reading of the draft. The  work of EC and AVR was supported in part by MICINN and  FEDER  under grant
FPA2008-01838,  by the Spanish Consolider-Ingenio 2010 Programme CPAN (CSD2007-00042) and by Xunta de Galicia (Conselleria de Educacion  grant INCITE09 206 121 PR). EC is supported by a spanish FPU fellowship. 
 \emph{Centro de F\'{\i}sica do Porto} is partially funded by the
PTDC/FIS/099293/2008 and the CERN/FP/109306/2009 programs.
DZ is funded by the FCT grant SFRH/BPD/62888/2009.

%%%%%%%%%%%%%%%%%%%%%%%%%%%%%%%%%%%%%%%%%%%%%%%%%%%%%%%%%%%%%%%%%%%%%%%%%%

\end{document}